%% file: AA_2018_33828.tex
\documentclass{aa}

\usepackage{graphicx}
\usepackage{txfonts}
\usepackage{amsmath}
\usepackage{amssymb}
\usepackage{textcomp}
\usepackage{siunitx}
\usepackage{fancyref}
\usepackage{color}
\usepackage{array}
\usepackage{natbib}
\usepackage{hyperref}
\bibpunct{(}{)}{;}{a}{}{,}

\begin{document}
 
 \title{Asymmetric jet production in the Active Galactic Nucleus of \object{NGC\,1052}}	
\author{%
  A.-K. Baczko\inst{\ref{affil:mpifr},\ref{affil:wuerzburg},\ref{affil:remeis}}
  \and R. Schulz\inst{\ref{affil:astron},\ref{affil:wuerzburg}}
  \and M. Kadler\inst{\ref{affil:wuerzburg}}
  \and E. Ros\inst{\ref{affil:mpifr}}
  \and M. Perucho\inst{\ref{affil:dep_valencia2},\ref{affil:dep_valencia}}
  \and C. M. Fromm\inst{\ref{affil:frankfurt},\ref{affil:mpifr}}
  \and J. Wilms\inst{\ref{affil:remeis}}
  }%
\institute{%
  Max-Planck-Institut f\"ur Radioastronomie, Auf dem H\"ugel 69, D-53121 Bonn, Germany\label{affil:mpifr}
  \and%
  Institut f\"ur Th. Physik und Astrophysik, Univ. W\"urzburg, Emil-Fischer-Str. 31, D-97074 W\"urzburg, Germany\label{affil:wuerzburg}
  \and%
Dr. Remeis-Sternwarte \& ECAP, Univ. Erlangen-N\"urnberg, Sternwartstr. 7, D-96049 Bamberg, Germany\label{affil:remeis}
  \and%
  Netherlands Institute for Radio Astronomy (ASTRON), PO Box 2, NL-7990 AA Dwingeloo, The Netherlands\label{affil:astron}
  \and%
  Observatori Astron\`omic, Univ. de Val\`encia, C/ Catedr\'atico Jos\'e Beltr\'an no. 2, E-46980 Paterna, Val\`encia, Spain
  \label{affil:dep_valencia2}
  \and%
  Dep. d'Astronomia i Astrof\'isica, Univ. de Val\`encia, C/ Dr. Moliner 50, E-46100 Burjassot, Val\`encia, Spain
  \label{affil:dep_valencia}
  \and%
  Institut f\"ur Theoretische Physik, Goethe Universit\"at, Max-von-Laue-Str. 1, 60438 Frankfurt, Germany\label{affil:frankfurt}
  }%
  
\abstract{
\textit{Context:} Few Active Galactic Nuclei (AGN) reveal a double sided jet system. However, these are crucial to understand basic physical properties of extragalactic jets. \newline
\textit{Aims:} We address the questions whether jets in AGN are symmetric in nature, how well they are collimated on small scales, and how do they evolve with time. \newline
\textit{Methods:} We monitored the sub-parsec scale morphology of \object{NGC\,1052} with the VLBA at 43\,GHz from 2005 to 2009.\newline
\textit{Results:} A detailed study of twenty-nine epochs show a remarkable asymmetry between both jets. 
A kinematic analysis of the outflows reveals higher apparent velocities for the eastern (approaching) jet as compared to the western (receding) one: $\beta_\mathrm{ej}=0.529\pm0.038$ and $\beta_\mathrm{wj}=0.343\pm0.037$, respectively. Contradictory to previous studies, we find higher flux densities for the western jet as compared to the eastern one. 
The distribution of brightness temperature and jet width features well collimated jets up to 1\,mas distance to the dynamic center and a nearly conical outflow further outwards. 
Combining flux density ratios and velocities of the jet flow, we were not able to find a combination of intrinsic velocity and inclination angle of the jets that is consistent for all four years of observation, which is contradictory to symmetrically evolving jets.
Spectral index maps between quasi-simultaneous 22\,GHz and 43\,GHz observations support the existence of an optically thick absorber covering the innermost $\simeq 1.6\,$mas around the 43\,GHz central feature and an optically thin jet emission with a spectral index of $\leq  -1$. \newline
\textit{Conclusions:} Our results fit into a picture in which we expect larger internal energy and/or magnetic flux in the western jet and higher kinetic energy in the eastern jet. Previous observations at lower frequencies found slower velocities  of the moving jet features as compared to this work. Considering the different velocities at different areas we suggest a spine-sheath structure with a faster inner layer and a slower outer layer.}

\maketitle

\section{Introduction}
\label{sec:intro}
Relativistic jets observed in radio-loud AGN are intensively studied, yet key questions remain regarding the formation, collimation and evolution especially on (sub-) parsec scales. Theoretical models by \cite{Bla79} or \cite{Koe81} describe narrow, conical evolving jets with a constant velocity. In addition, most models assume two symmetric jets ejected in opposite directions from the central engine. Numerical simulations, however, show that asymmetric jets can be produced \citep[see e.g.,][]{Fen13}. Observationally, it is difficult to find a suitable AGN with (1) both jets prominently visible and sufficiently strong at high frequencies, (2) the jets oriented in the plane of the sky to avoid effects of the differential Doppler boosting, and (3) a close distance in order to reach small spatial scales.
 
The radio galaxy \object{NGC\,1052} fulfills these requirements. It hosts a super massive black hole (SMBH) with a mass of the order of $\approx10^{8.2}\,\mathrm{M_\odot}$ \citep{Woo02}.
Due to its optical spectrum, it is classified as a low-ionization nuclear emission-line region (LINER) object \citep{mayall:1939,fosbury:1978,ho97}. 

\object{NGC\,1052} is located at a redshift of $z=0.005037$ (assuming $v_{sys}=1507\mathrm{km\,s}^{-1}$, \citealt{Jen03}). For nearby objects such as \object{NGC\,1052} distance estimates based on recession velocities are biased by peculiar velocities of the galaxies. We use, therefore, the redshift-independent distance of $D=19.23\pm 0.14\,$Mpc \citep{Tul13}. This results in a linear scale of $0.093\,\mathrm{pc/mas}$.

At pc-scales \object{NGC\,1052} exhibits two jets, which emanate from the center, and show an emission gap at cm-wavelengths that becomes smaller towards higher frequencies. This effect is due to free-free absorption caused by a geometrically thick structure perpendicular to the jets with an optical depth $\tau_\mathrm{1\,GHz}\sim 300$ to $1000$ \citep{Kameno2001,sawada-satoh2008} and a column density of $10^{22}\,\mathrm{cm^{-2}}$ to $10^{24}\,\mathrm{cm^{-2}}$ \citep{Kameno2001,Vermeulen2003,Kad04a}. This circumnuclear torus covers larger parts of the western, receding jet. Its absorbing effect becomes smaller towards higher frequencies, and has close to no impact at 43\,GHz.
Several authors reported on $\mathrm{H_2O}$ maser emission, which is most likely associated with the surrounding torus \citep{cla98,kam05,sawada-satoh2008}. Based on multi-frequency observations 
performed in 
2000 \cite{sawada-satoh2008} identified the $\mathrm{H_2O}$ masers projected on the eastern and western jet components B and C3, respectively.
The mean speed of features in both jets have been derived $v/c=\beta\leqslant 0.23$ \citep{Vermeulen2003,boeck2012,Lis13}.
 
\citet{Baczko16} presented the first detection of the twin jets in \object{NGC\,1052} at 86\,GHz. These results show a compact, bright central feature and two faint plasma streams. The central feature can be interpreted as blended emission from both jet cores. Therefore, this central feature pinpoints the location of the central engine. Assuming basic synchrotron theory, this sets boundaries to the strength of the magnetic field at $1\,R_\mathrm{S}$: $200\,\mathrm{G}<B<8\times 10^4\,\mathrm{G}$.

In this paper, we investigate this exceptional source further at 43\,GHz and 22\,GHz based on twenty-nine observations over four years with the Very Long Baseline Array (VLBA).
In Section~2 we introduce the observations. Section~3 gives an overview on the methods used to derive the physical quantities. In Section~4, we discuss the different aspects of our analysis, and Section~5 summarizes our results and gives an outlook. 
 
\section{Observation and Data Reduction}
\label{sec:Observation_Reduction}
 \subsection{Imaging}
 \label{sec:Maps}
\object{NGC\,1052} was observed with the VLBA over twenty-nine epochs (full track) between 2005 March and 2009 April quasi simultaneously at 22\,GHz and 43\,GHz. The data were recorded with eight sub-bands with a bandwidth of 8\,MHz and data rate of 256\,Mbps. The fringe finding calibrators were \object{III\,Zw\,2} and \object{B0420--014}. The correlation was performed at the National Radio Astronomy Observatory (NRAO) at the Array Operation Center of the VLBA in Socorro, NM, USA.
  
A priori amplitude calibration was performed using antenna gain curves and measured system temperatures at all stations. Solutions have been found for the relative delays between observing sub-bands (manual phase-cal) and a fringe search has been conducted to obtain residual delays and delay rates (fringe process). These tasks were performed in AIPS. After averaging over frequencies within each sub-band data were exported from \textsc{aips} to perform hybrid mapping in \textsc{difmap} \citep{Shepherd1994}. Several iterations of the \textsc{clean} algorithm and phase and amplitude self-calibration loops yielded the final images used in our analysis. Table~\ref{tab:map_pars} lists the final $43\,$GHz image parameters.

  \begin{table*}
   \setlength{\tabcolsep}{2pt}
   \small
   \centering
   \caption{Image parameters for all analyzed observations at 22\,GHz and 43\,GHz with natural weighting.}
   \label{tab:map_pars}
   \begin{tabular}{lllcccccc}\hline\hline
  Epoch &VLBA code & Frequency & RMS & $S_\mathrm{peak}$&$S_\mathrm{tot}$&$b_\mathrm{maj}$&$b_\mathrm{min}$& PA \\
	&    & $[$GHz$]$ & $[\mathrm{\frac{mJy}{beam}}]$&$[\mathrm{\frac{Jy}{beam}}]$&[Jy]&[mas]&[mas]&[$^\circ$]\\
	(1) & (2) & (3) & (4) & (5) & (6) & (7) & (8) & (9)\\\hline
2005-03-14 $^{\dag}$						& BR099A	& 43 &	1.21	&	0.33	&	0.80	&	0.46	&	0.20	&	$7.37$	 \\
&& 22 & 8.33 &  0.27 &0.84 &0.72 & 0.30 & $4.64$ 	\\
2005-04-22 $^{\star}$ $^{\dag}$ & BR009B	&	43 & 	1.23	&	0.29	&	1.64	&	1.11	&	0.21	&	$-18.56$ \\
&& 22 & 0.70 &  0.53 &2.09 &2.32 & 0.36 & $-18.58$ \\
2005-06-06 											& BR099C	&	43 & 	0.99	&	0.41	&	1.17	&	0.69	&	0.20	&	$-16.85$ \\
&& 22 & 0.68 &  0.64 &2.26 &0.96 & 0.37 & $-10.34$\\
2005-07-18 											& BR099D	&	43 & 	0.75	&	0.52	&	1.00	&	0.52	&	0.18	&	$-11.36$ \\
&& 22 &  0.50 &  0.21 &1.08 &0.80 & 0.31 & $-4.86$ 	\\
2005-08-22 											& BR099E	&	43 & 1.20	&	0.47	&	1.51	&	0.51	&	0.18	&	$-12.80$ \\
&& 22 &  0.85 &  0.58 &1.99 &0.89 & 0.32 & $-9.62$ \\
2005-10-07 											& BR099F	&	43 & 1.05	&	0.48	&	1.30	&	0.49	&	0.19	&	$-8.47$	 \\
&& 22 &  0.75 &  0.44 &1.39 &0.86 & 0.33 & $-6.25$ \\
2005-11-13 											& BR099G	&	43 & 0.68	&	0.26	&	0.80	&	0.50	&	0.18	&	$-9.88$	 \\
&& 22 &  0.31 &  0.16 &0.95 &0.92 & 0.33 & $-9.04$ \\
2005-12-18 $^{\star}$ $^{\dag}$	& BR009H	&	43 & 1.02	&	0.40	&	0.91	&	0.64	&	0.20	&	$-11.64$ \\
&& 22 &  0.55 &  0.40 &0.94 &1.09 & 0.34 & $-10.05$ \\
2006-01-15 											& BR099I	&	43 & 0.84	&	0.32	&	0.78	&	0.51	&	0.20	&	$-8.11$	 \\
&& 22 &  2.00 &  0.29 &1.30 &0.70 & 0.26 & $-7.23$ 		\\
2006-03-06 											& BR119A	&	43 & 0.93 	&	0.40	&	0.83	&	0.52	&	0.20	&	$-8.12$	 \\
&& 22 &  5.58 &  0.36 &1.28 &0.65 & 0.26 & $-2.84$ \\
2006-04-19 $^{\dag}$						& BR119B	&	43 & 0.74	&	0.41	&	1.04	&	0.48	&	0.18 	&	$-7.15$	 \\
&& 22 &  1.69 &  0.19 &1.23 &0.65 & 0.25 & $-4.03$ \\
2007-01-14 $^{\dag}$						& BR120A	&	43 & 0.41	&	0.21	&	0.74	&	0.61	&	0.26	&	$10.05$	 \\
&& 22 &  0.30 &  0.21 &0.61 &0.86 & 0.36 & $-3.54$ 	\\
2007-03-04 											& BR120B	&	43 & 0.35	&	0.24	&	0.69	&	0.41	&	0.17	&	$-3.73$	 \\
&& 22 &  0.27 &  0.23 &0.67 &0.89 & 0.36 & $-2.42$ \\
2007-05-05 											& BR120C	&	43 & 0.42	&	0.25	&	0.49	&	0.40	&	0.16	&	$-5.15$	 \\
&& 22 &  0.35 &  0.26 &0.59 &0.78 & 0.31 & $-5.11$ \\
2007-06-17 											& BR120D	&	43 & 0.38	&	0.30	&	0.64	&	0.51	&	0.17	&	$-8.27$	 \\
&& 22 &  0.32 &  0.28 &0.70 &0.93 & 0.32 & $-5.73$ 	\\
2007-08-11 											& BR120E	&	43 & 0.44	&	0.26	&	0.56	&	0.53	&	0.18	&	$-12.30$ \\
&& 22 &  0.32 &  0.27 &0.63 &0.84 & 0.33 & $-4.99$ \\
2007-10-01 $^{\dag}$						& BR120F	&	43 & 0.40	&	0.29	&	0.63	&	0.88	&	0.49	& $21.76$	 \\
&& 22 &  0.30 &  0.29 &0.60 &1.40 & 0.39 & $-16.64$ 	\\
2008-01-11 $^{\star}$ $^{\dag}$ & BR120G	&	43 & 0.67	&	0.28	&	0.66	& 0.59	&	0.20	&	$-11.12$ \\
&& 22 &  0.17 &  0.22 &0.50 &0.96 & 0.34 & $-5.99$ 	\\
2008-04-04 											& BR120H	&	43 & 0.37	&	0.21	&	0.73	&	0.54	&	0.19	&	$-11.98$ \\
&& 22 &  0.19 &  0.29 &0.72 &1.22 & 0.37 & $-15.58$\\
2008-05-27 											& BR120I	&	43 & 0.77	&	0.27	&	0.58	&	0.48	&	0.20	&	$-4.82$	 \\
&& 22 &  0.21 &  0.24 &0.64 &0.92 & 0.33 & $-8.59$ \\
2008-05-04 											& BR130A	&	43 & 0.56	&	0.30	&	0.70	&	0.42	&	0.16	&	$-6.07$	 \\
&& 22 &  0.24 &  0.25 &0.72 &0.87 & 0.31 & $-8.35$ \\
2008-06-16 											& BR130B	&	43 & 0.55	&	0.27	&	0.63	&	0.50	&	0.21	&	$-2.34$	 \\
&& 22 &  0.24 &  0.24 &0.65 &0.86 & 0.33 & $-6.65$ \\
2008-07-26 											& BR130C	&	43 & 0.70	&	0.36	&	0.64	&	0.47	&	0.20	&	$-4.37$	 \\
&& 22 &  0.31 &  0.20 &0.52 &0.77 & 0.29 & $-6.67$ 	\\
2008-09-06 											& BR130D	&	43 & 0.77	&	0.31	&	0.69	&	0.59	&	0.25	&	$-5.82$	 \\
&& 22 &  0.27 &  0.16 &0.67 &0.95 & 0.32 & $-10.85$\\
2008-10-27 $^{\dag}$						& BR130E	&	43 & 0.47	&	0.11	&	0.78	&	0.44	&	0.18	&	$-4.77$	 \\
&& 22 &  0.24 &  0.21 &0.95 &0.89 & 0.32 & $-8.94$ \\
2008-12-05 $^{\dag}$						& BR130F	&	43 & 0.50	&	0.22	&	0.72	&	0.76	&	0.32	&	$15.87$	 \\
&& 22 &  0.20 &  0.17 &0.86 &0.90 & 0.33 & $-7.99$ 	\\
2009-01-18 $^{\dag}$						& BR130G	&	43 & 0.50	&	0.26	&	0.69	&	0.99	&	0.37	&	$22.77$	 \\
&& 22 &  & & & & &\\ 
2009-03-08 											& BR130H	&	43 & 0.41	&	0.15	&	0.64	&	0.44	&	0.18	&	$-3.25$	\\
&& 22 &  0.26 & 0.18 & 0.71 &0.91 & 0.34 & $-4.37$ \\
2009-04-16 											& BR130I	&	43 & 0.32	&	0.16	&	0.65	&	0.43	&	0.16	&	$-7.43$	\\
&& 22 &  0.21 & 0.26 & 0.82 &0.90 & 0.34 & $-6.9$ 	\\
   \hline
   \end{tabular}
    \tablefoot{
	(1) date of VLBA observation $^{\star}$: excluded in 43\,GHz analysis $^{\dag}$: excluded in combined 43\,GHz and 22\,GHz analysis,
	(2) VLBA experiment code,
	(3) Frequency
	(4) rms noise level of image. RMS was derived in \textsc{ISIS} by fitting a Gaussian profile to the distribution of pixel values in the images, the rms corresponds to the sigma of the best fit profile., 
	(5) peak flux density,
	(6) total flux density,
	(7) FWHM major axis of restoring beam,
	(8) FWHM minor axis of restoring beam,
	(9) position angle of major axis of restoring beam.
 }
  \end{table*}
 
\subsection{Gaussian model fitting}
 \label{sec:Modelfits}

To parameterize the image properties, we fitted two-dimensional Gaussian functions to the visibility data in \textsc{difmap}. This was performed independently for each observation until satisfactory models for all epochs were obtained. Hence, the model is stable and reduces the noise in the image as much as possible. The parameters of all model fitting procedures are listed in the Appendix (Tables \ref{tab:timeevo_fits_single1} to \ref{tab:timeevo_fits_single5}). 
The parameters `total flux', 'distance' and 'major axis' had been set free in the model. The axial ratio for most jet components had been fixed to unity to get the most satisfactory model, whereas it was left as free parameter for a large fraction of the core components. 
For component identification, images were aligned on the central bright feature that is observable in almost all epochs. It is located at the map peak, and, therefore, at the map origin. This approach differs from that used by \cite{boeck2012}, who aligned the maps at 15\,GHz, 22\,GHz and 43\,GHz on the center of the emission gap, that is dominant at frequencies of 22\,GHz and below. As we see no indication of strong absorption effects at 43\,GHz, the central bright feature is assumed to pinpoint the position of the central supermassive black hole and thus remains constant over time. The models were iterated, meaning different combinations of number, position, and shape of Gaussian components were tested, to find consistency between adjacent epochs. Three datasets were excluded from the analysis (BR099B, BR099H, and BR120G) due to poor data quality. For the amplitude values we assumed an uncertainty of 10\%, which represents the typical uncertainity for the amplitude in case of the VLBA. Based on our sample of 43\,GHz observations we estimated the uncertainty of amplitude by means of gscale statistics in difmap to be slightly higher with an uncertainty of 14\%. We derived the standard deviation of gscale values for all IFs and antennas assuming a `true' mean value of unity. The uncertainty of amplitude is then given by the median of the final distribution for all observations. However, this analysis includes observations which turned out to be very noisy and such that are very accurate. In the case of the latter the uncertainty would be largely overestimated. Therefore, we stick to the previously mentioned 10\% uncertainty of amplitude.

To perform a linear regression for deriving speeds, we assigned uncertainties to the data points as follows: the error bar in the width of each model-fit component $m$ was set to $\sigma_{W_m} = W_m/(S/N)_{m}$, where $(S/N)_{m}$ is the signal-to-noise ratio $(S/N)$ of the component peak to the residual noise in the model-fit map, and $W_m$ is the full-width half maximum of the Gaussian model-fit component \citep{Fom99}.
The error on the position of the Gaussian components depends on the remaining noise in the \textsc{clean} and model-fit maps, the size of the components and the beam axis. 
Therefore, the positional uncertainty was defined as:
\begin{equation}
    \mathrm{\delta}\theta=\sqrt{\left(\frac{b_\mathrm{paj}}{(S/N)_{c}}\right)^2+\left(\frac{a_\mathrm{paj}}{(S/N)_{m}}\right)^2},
    \label{equ:pos_err}
\end{equation}
with $b_\mathrm{paj}$ is the beam axis and $a_\mathrm{paj}$ is the component axis along the jet position angle (P.A.) which is equal to $64^\circ$. $(S/N)_{c/m}$ is the signal-to-noise ratio of the map peak to the residual noise in the \textsc{clean}/model-fit map. 
In the case of very small errors for small component sizes we set a lower boundary to the positional error equal to the beam size divided by 10, following the approach by \citet{Lis09}.

\section{Results}
\label{sec:Analysis}

 There are two periods without observations of about half a year. Over these periods the morphology of the source changes significantly. Hence, we divide the data into three time blocks:  ``Block~1'': from 2005 to 2006; ``Block~2'':  during 2007 and ``Block~3'': from 2008 to 2009. The individual images, with the Gaussian model fit components plotted on top of the \textsc{clean} maps are shown in Figures \ref{fig:TimeEvo1} and \ref{fig:TimeEvo2}.

\subsection{Stacking}
    
\begin{figure}[!h]
  \centering
    \includegraphics[width=0.9\linewidth]{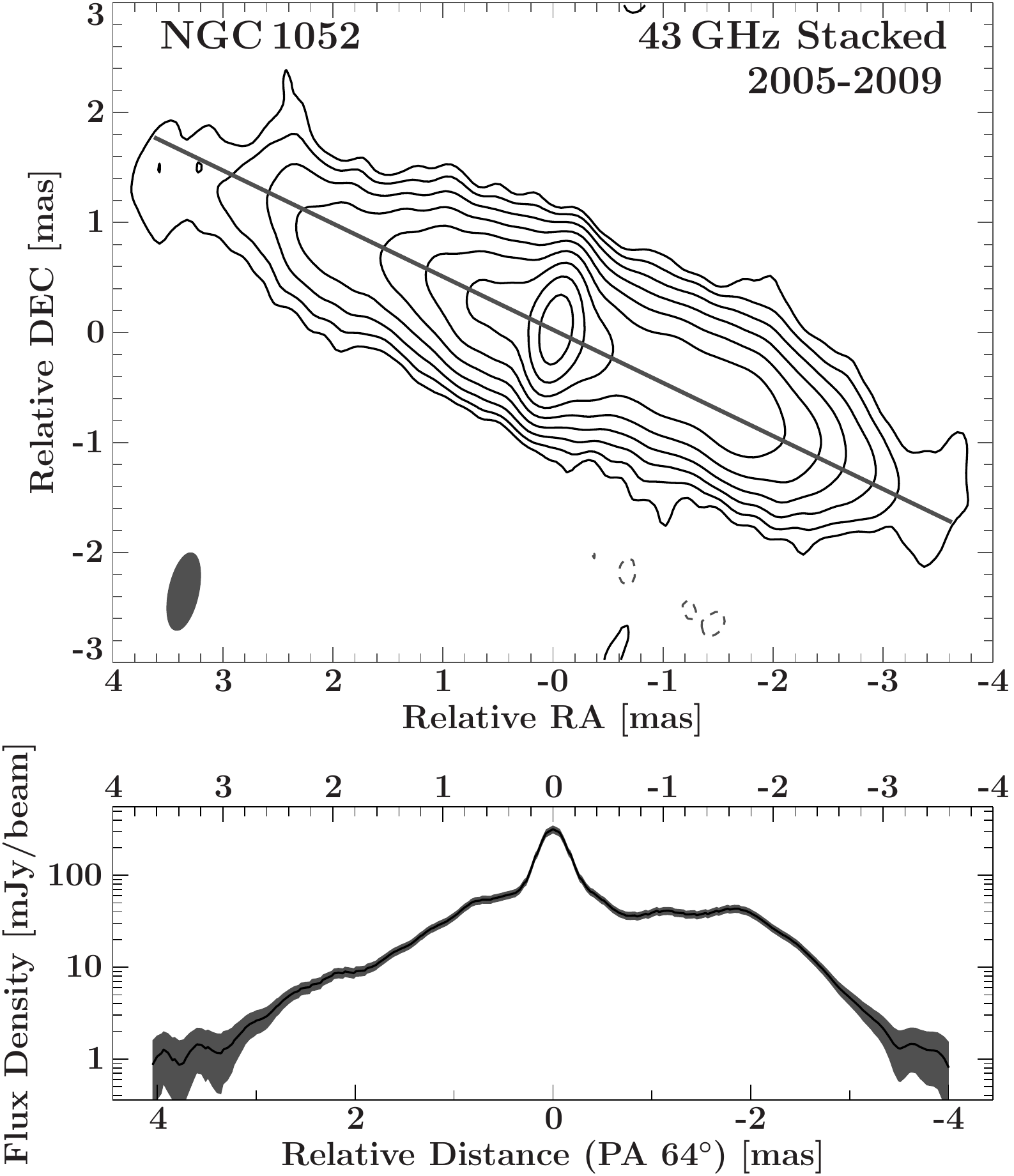}\hspace{0.04\linewidth}
    \caption{Stacked image for all observations from 2005 to 2009 (\textit{top panel}) and flux density along a cut in the jet direction of $64\si{\degree}$ (\textit{bottom panel}). This cut is indicated as a line along the jet axis in the contour map. The common beam for the stacked map is show in the lower left corner of the \textit{top panel}. For the stacked image, all maps have been summed up and divided by the number of images. The contours start at 5 times the noise level and increase logarithmically by factors of 2.}
    \label{fig:RidgeLine}
\end{figure}

After aligning all images to the bright central peak, we produced a stacked image by adding pixel-by-pixel all analyzed 43\,GHz observations followed by averaging (see Fig.~\ref{fig:RidgeLine}) similar to the procedure described in \cite{Pus17}. 
Both jets are fairly symmetric and have an extent of around $4\,$mas at this wavelength. 
A cut along the jet position angle at $64^\circ$ (from north to east) gives the flux density profile along the jet axes and is illustrated in the lower panel. It forms a plateau from $-1$\,mas to $-2$\,mas along the cut, and after which the flux density of the western jet drops off more steeply than the eastern jet.
  
 \subsection{Spectral index maps between 22\,GHz and 43\,GHz}
 \label{sec:Spec_align}
We make use of simultaneous 22\,GHz observations to study the spectral index distribution of the twin-jet system between 43\,GHz and 22\,GHz (Section~\ref{sec:Spectral_analysis}). Prior to calculating the spectral indices, the images at both frequencies were aligned (Section~\ref{sec:alignment}) and convolved with a common beam. Four observations were found to have insufficient image quality for a proper alignment and further five observations did not have a reliable Gaussian modelfit at 22\,GHz and thus were excluded. After the exclusion of these epochs we were able to obtain nineteen epochs which remain available for a reliable spectral index analysis.

  \subsubsection{Alignment}
  \label{sec:alignment}
  
\begin{figure*}
    \centering
    \includegraphics[width=0.9\linewidth]{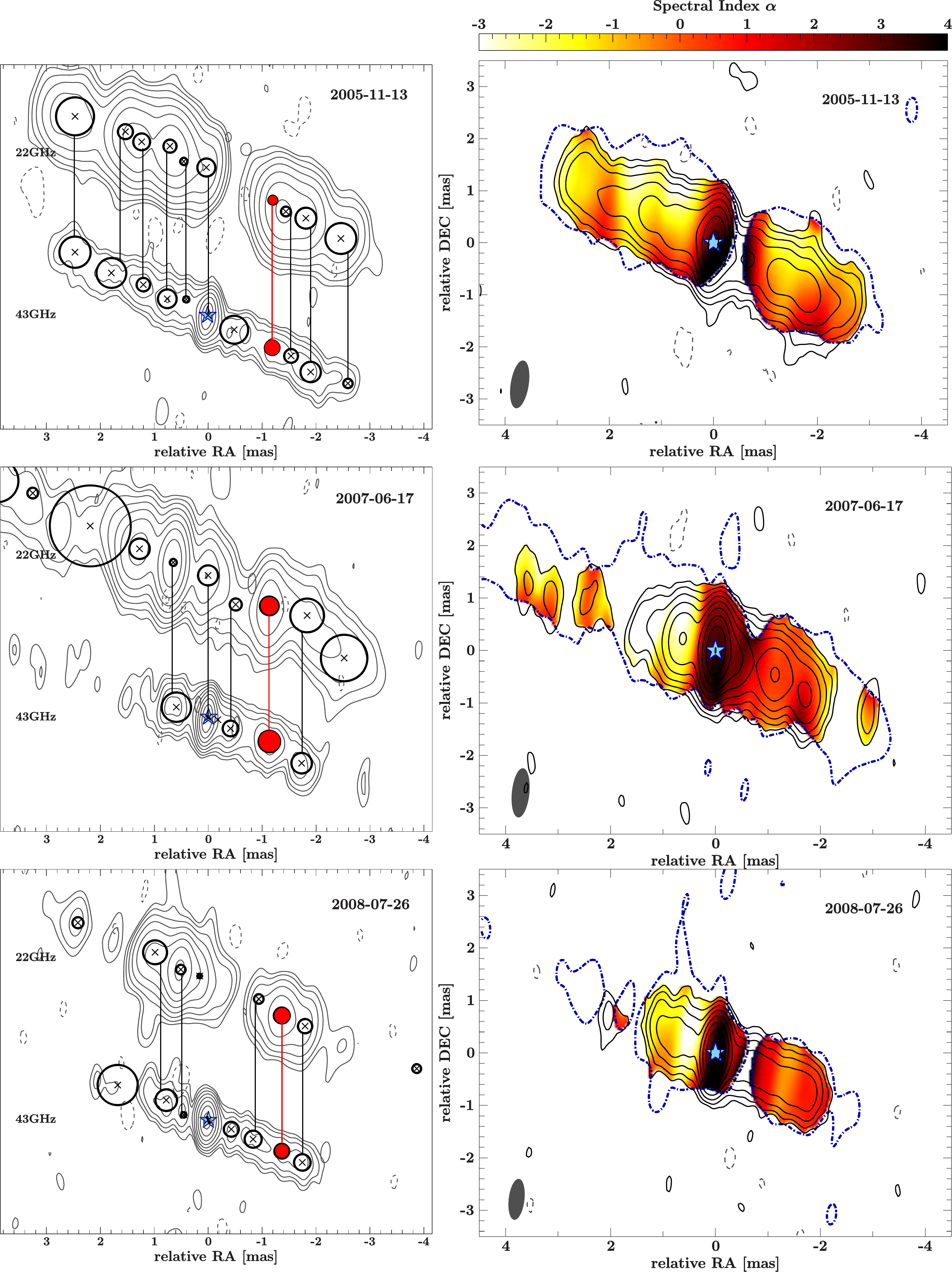}
    \caption{
    (\textit{left}) Three examples of the alignment of the 22\,GHz and 43\,GHz maps. We used the optically thin feature $\sim 2\,$mas to the west of the 22\,GHz jet core (highlighted in red) to derive the relative shift between the two frequencies. 
    The Gaussian model components are plotted on top of the contour clean maps. 
    Vertical lines connect the components identified between both frequencies. 
    The contours start at 3 times the noise level and increase logarithmically by factors of 2. 
    The beam parameters for the six observations are listed in Tab.~\ref{tab:map_pars}.
    (\textit{right}) Resulting spectral index maps between 22\,GHz (dotted dashed blue contours) and 43\,GHz (black contours) after applying the shifts on optically thin features. 
    The restoring beam for each epoch is plotted in the lower left corner of the maps, parameters are listed in Tab.\ref{tab:spix_examples}. The blue star marks the assumed location of the dynamical center.
    }
    \label{fig:spix_blocks_examples}
\end{figure*}

We used optically thin features in both jets to align the maps at both frequencies \citep{Kad04b,Fro13}. The most suitable feature for the alignment is located at around $-2\,$mas in RA distance to the eastern jet core in the 22\,GHz map. 
Figure~\ref{fig:spix_blocks_examples} gives an example of the alignment applied for the beginning, middle, and the end of the observational campaign. These three maps are a good representation of the typical morphology during the individual time blocks and reveal easy to compare Gaussian model fit components.
In case of the 22\,GHz images we used the Gaussian model fits produced by \cite{boeck2012}, which are distributed similarly to the components at 43\,GHz. For the alignment the beam parameters have not been changed (see Tab.~\ref{tab:map_pars} for the individual parameters).
   
The alignment of the maps sets the central peak in the 43\,GHz map around $0.6\,$mas to the west of the 22\,GHz eastern jet peak. For all three examples, the 22\,GHz model reveals one Gaussian component at the position of the 43\,GHz central component. 
This is the case for a large fraction of the 19 epochs used here although the 22\,GHz central component is substantially fainter and less well defined then the corresponding 43\,GHz component. 

The four years of observation reveal a rapidly changing twin-jet system. This makes it difficult to compare with earlier observations. For example, in a multi-frequency observation in 2000, \cite{sawada-satoh2008} identified the brightest feature at 43\,GHz with the eastern jet core at 22\,GHz. A nearby western component, identified as C4, resides close to the location where the central engine is suspected. Accounting the difference in flux density to the variability of the jets, it may well be identified with the central feature from our VLBA images.
   
  \subsubsection{Spectral analysis}
  \label{sec:Spectral_analysis}

After aligning the epochs as discussed in Section~\ref{sec:alignment}, spectral index maps are produced based on the hybrid images. Representative spectral images are shown in the right column of  Fig.~\ref{fig:spix_blocks_examples} and the common beams for both frequencies used for each map are listed in Tab.~\ref{tab:spix_examples}. The spectral index distributions reveal significant absorption in the region of the gap at 22\,GHz, reaching spectral indices of $\alpha=4$ in individual maps. This supports previous results of a free-free absorbing torus covering parts of the western jet \citep{Kameno2001,sawada-satoh2008} up to 43\,GHz \citep{Vermeulen2003,Kad04b}. 
   
\begin{table}[!htb]
 \setlength{\extrarowheight}{1.5pt}
 \setlength{\tabcolsep}{3pt}
 \small
 \centering
 \caption{Parameters used for the common beam in the spectral index maps shown in Fig.~\ref{fig:spix_blocks_examples}}
 \label{tab:spix_examples}
 \begin{tabular}{lcccc}\hline\hline
     Epoch & $b_\mathrm{maj}$&$b_\mathrm{min}$& PA \\
	& [mas]&[mas]&[$^\circ$]\\
	(1) & (2) & (3) & (4)\\\hline
     2005-11-13 & 0.92 & 0.33 & $-9,04$ \\
     2007-06-17 & 0.93 & 0.32 & $-5.73$ \\
     2008-07-26 & 0.77 & 0.29 & $-6.67$ \\
     \hline
 \end{tabular}
 \tablefoot{
	(1) date of VLBA observation,
	(2) FWHM major axis of restoring beam,
	(3) FWHM minor axis of restoring beam,
	(4) position angle of major axis of restoring beam
    }
\end{table}
      
\begin{figure}[!h]
    \centering
    \includegraphics[width=0.95\linewidth]{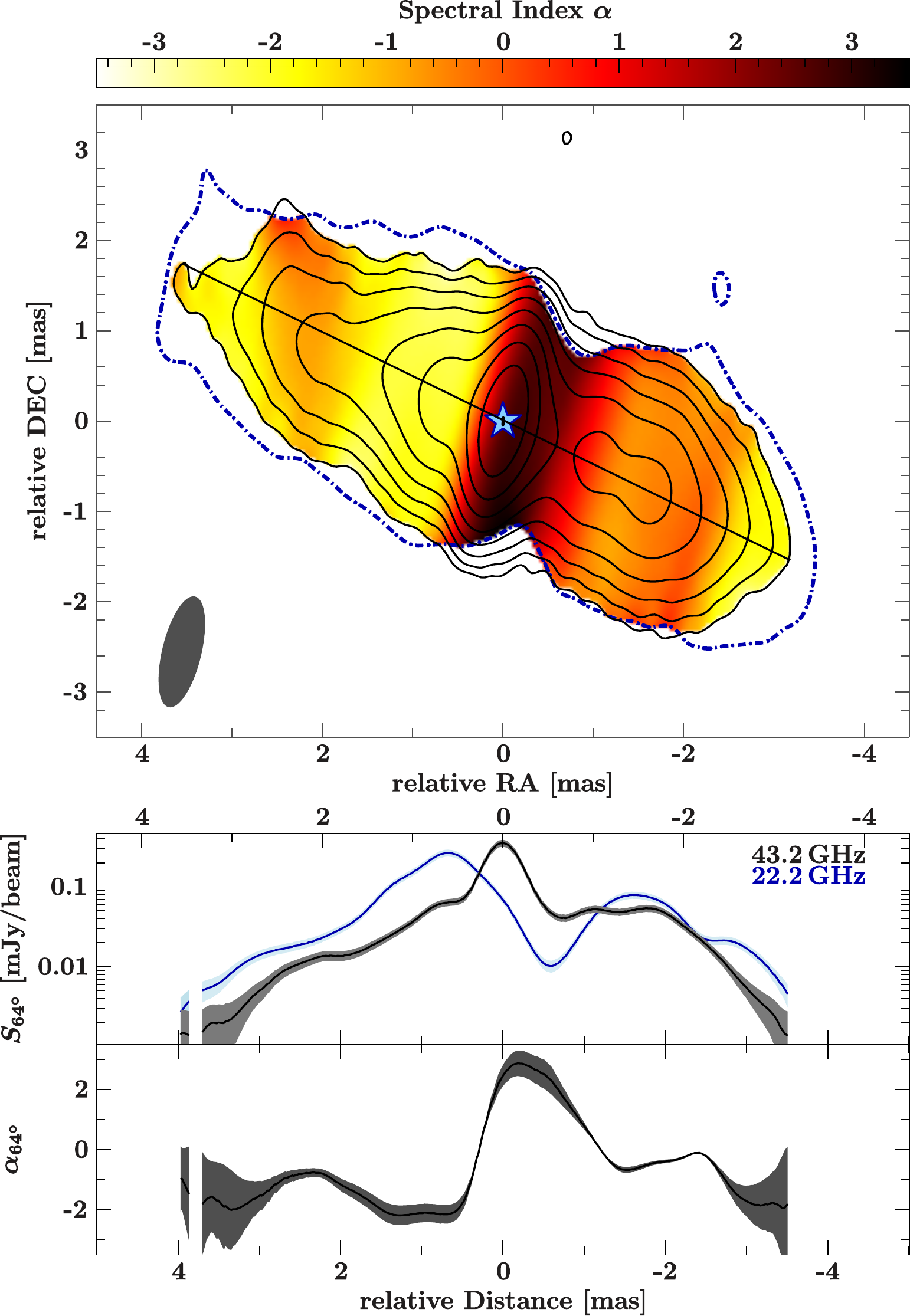}
    \caption{ Stacked spectral index map ($S \propto \nu^{+\alpha}$) combining nineteen 22\,GHz (blue dashed-dotted contour) and 43\,GHz (black contours) images between 2005 and 2009. 
	A cut along the direction of the jets is shown in the lower panels, depicting the flux density profile and spectral index along the jet axis (see line in upper panel). The blue star marks the assumed location of the dynamical center.
    The contours of the 43\,GHz map start at 5 times the noise level and increase logarithmically by factors of 2. For the 22\,GHz map only one countour at 5 times the noise level is shown.}
    \label{fig:sm_stacked}
\end{figure}

Figure \ref{fig:sm_stacked} shows a spectral-index map derived from one stacked map for all 19 epochs at both frequencies. 
Before stacking, the images were restored with a common beam and images were aligned as discussed in the previous section.
      
 \subsection{Flux density and kinematics}
 \label{sec:Flux_Kinematics}
  \subsubsection{Flux density evolution}
   \label{sec:Flux}

\begin{figure}[!ht]
     \centering
     \includegraphics[width=0.95\linewidth]{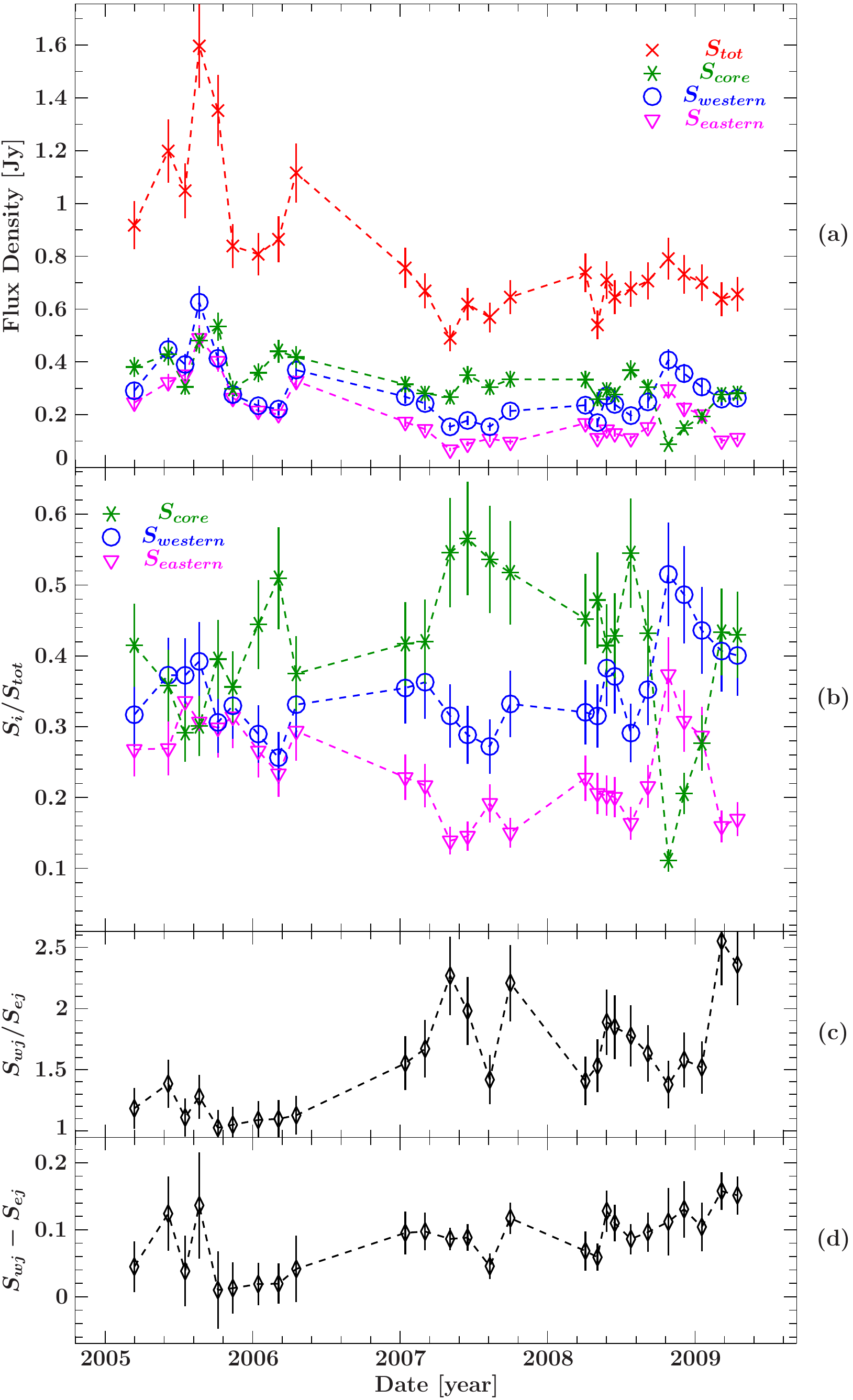}
     \caption{(a) Lightcurve at 43\,GHz of the total flux density (red), the core flux density (green), the western jet flux density (blue), and the eastern jet flux density (pink); (b)fractional intenities (relative to total) of the core component (green), the western jet (blue), and the eastern jet (pink); (b) fraction of the core, eastern and western jet flux densities to the total flux density; (c) ratio of the western to the eastern jet flux densities; (d) absolute difference of the flux density of the western and the eastern jet.}%
     \label{fig:lightcurve}
    \end{figure}

    The evolution of the flux density over time is tracked by summing up the Gaussian model components of the core as well as components of the western and eastern jets. 
    The evolution of the total flux density and the flux density for each region at 43\,GHz is plotted in panels (a) and (b) of Fig.~\ref{fig:lightcurve}. 
    \object{NGC\,1052} shows significant variability over the four years of our observations. The total flux density rises to a peak of 1.6\,Jy in 2005 August and then declines below 0.8\,Jy during 2007 April and remains at this level for the rest of the observations. 
   
    The flux density evolutions of the core as well as the western and eastern jets follow the trend of the total intensity evolution. 
    In contrast to previous observations in the years 1995-2002, and 2004 \citep{Vermeulen2003,Kad04b,Baczko16}, the western jet is brighter than the eastern one (see (c) in Fig.~\ref{fig:lightcurve}) for the whole period 2005-2009. Due to this deviation we assumed the western jet to approach and the eastern jet to recede.

\begin{figure}[!ht]
     \centering
     \includegraphics[width=0.98\linewidth]{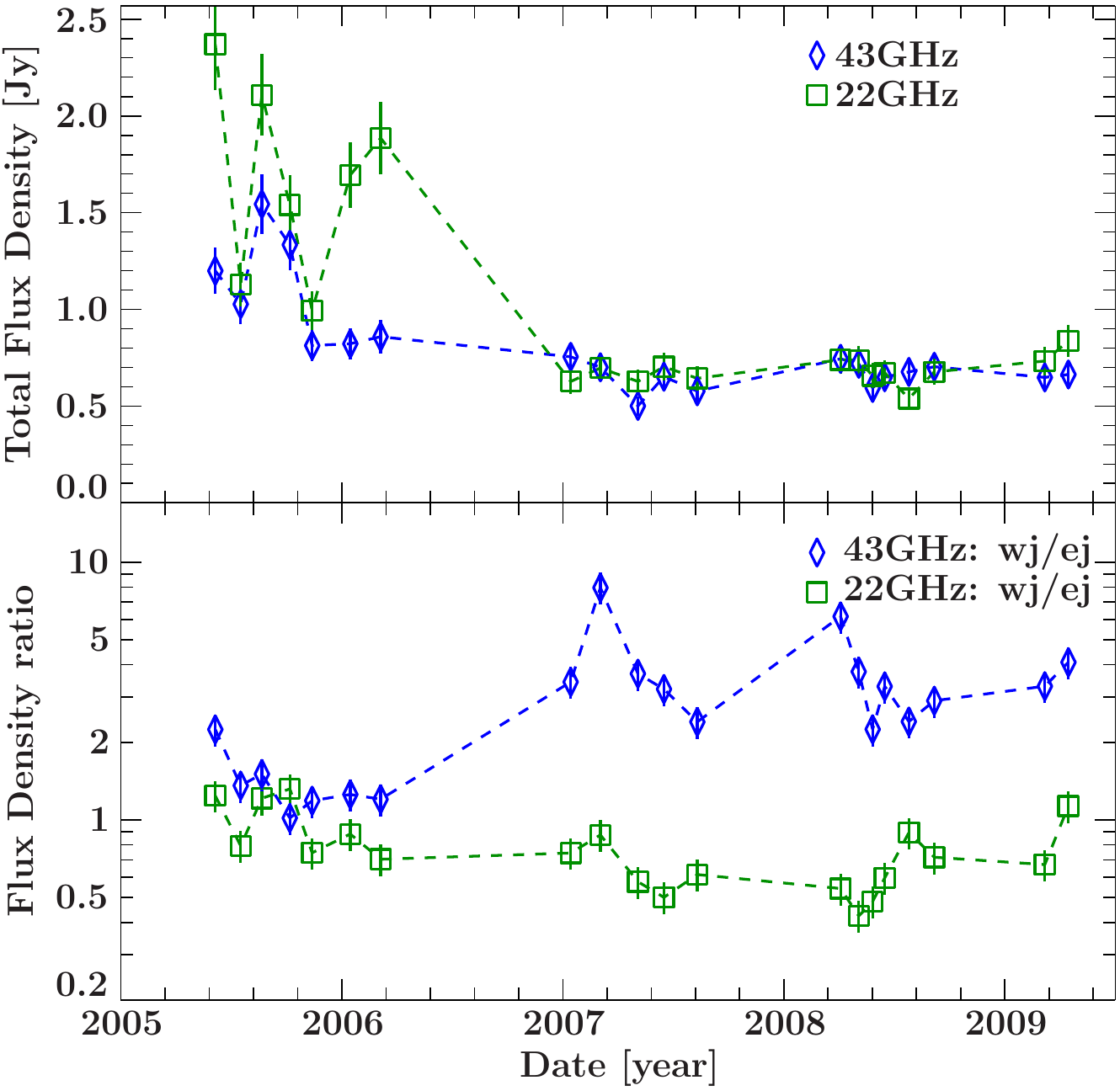}
     \caption{(\textit{top:}) total flux density of 19 clean maps at 22\,GHz (green squares) and 43\,GHz (blue diamonds). (\textit{bottom:}) flux density ratio of western to eastern jet at 22\,GHz (green squares) and 43\,GHz (blue diamonds).}
     \label{fig:flux_43_22GHz}
    \end{figure}

    We compared our results gained from the analysis of the Gaussian model fits to the initial \textsc{clean} images at 43\,GHz and 22\,GHz by deriving the flux density evolution based on the distribution of positive \textsc{clean} components in the images at 43\,GHz and 22\,GHz. Nineteen observations have been included in this comparison, analogous to the spectral analysis in Section~\ref{sec:Spec_align}. The total flux density and ratio between the brighter and the fainter jet is shown in Fig.~\ref{fig:flux_43_22GHz}. Similar to  Section~\ref{sec:Spectral_analysis} we first apply the shifts between the 22\,GHz and 43\,GHz maps, before excluding \textsc{clean} components up to 0.8\,mas distance in East and West to the 43\,GHz core at both frequencies to include only the optically thin jet regions. In this way we make sure, that we only include the parts of the jets at both frequencies, that are, indeed, comparable. All remaining \textsc{clean} components have been summed up for the eastern and the western jet. 

    The comparison of Fig.~\ref{fig:lightcurve} and Fig.~\ref{fig:flux_43_22GHz} shows that both models (Gaussian and \textsc{clean}) result in a consistent evolution of the flux density with time. However, without the need to exclude epochs due to insufficient resolution at 22\,GHz Fig.~\ref{fig:lightcurve} has a more dense time sampling. In contrast to 43\,GHz we only find 4 epochs at 22\,GHz in which the western jet is brighter than the eastern jet (2005-06-06, 2005-08-22, 2005-10-07 and 2009-04-16).

  \subsubsection{Kinematic analysis}
    \label{sec:Kinematics}

\begin{figure*}[!ht]
     \centering
     \includegraphics[width=0.75\linewidth]{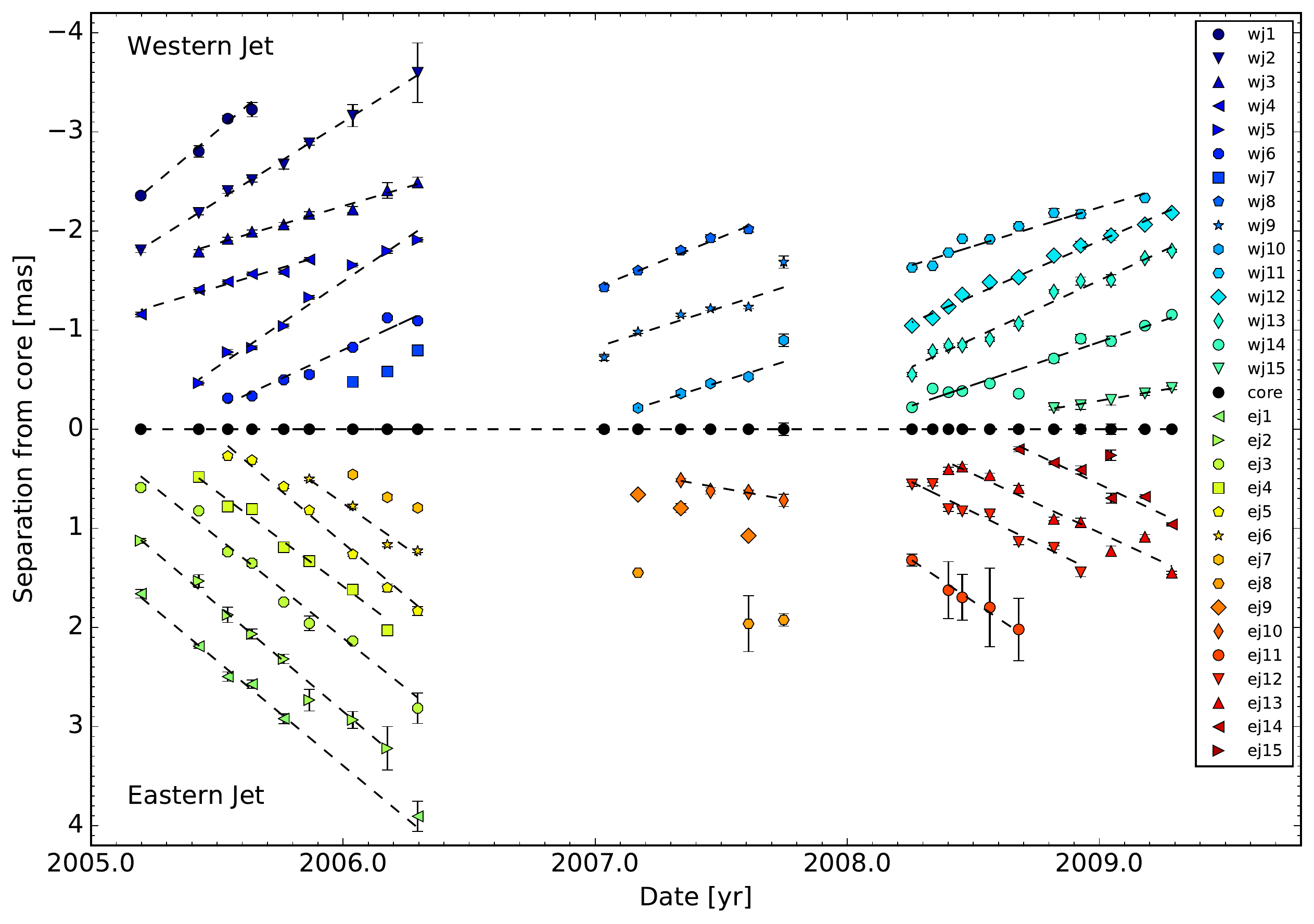}
     \caption{Core separation of individual Gaussian model-fit components over time. The fit to a linear function for each component is indicated by a dashed line (corresponding slopes are listed in Tab.~\ref{tab:speeds}). The eastern jet is plotted in the lower half of the plot, the western jet in the upper half of the plot.}
     \label{fig:timeevofit}
    \end{figure*}

    The Gaussian model components in the twenty-six 43\,GHz maps have been used to examine the dynamics in both jets. The central peak, observable at all epochs at 43\,GHz, has been assumed to be the dynamical center of the source. This is a natural choice based on the clear morphology at 43\,GHz and 86\,GHz \citep[compare Fig.~4 in][]{Ba16b}. In the unlikely case, that the 43\,GHz core is offset from the dynamical center, we still would expect that the 43\,GHz core does not shift with time. The same holds for the dynamical center and therefore the results obtained in this section are not affected.
The Gaussian component fitted to this feature was therefore used to align the maps. The Gaussian model components were cross-identified between adjacent observing epochs. This enabled us to apply linear fits to this feature's core distance as a function of time (see Fig.~\ref{fig:timeevofit}). The jet system is rapidly changing, therefore, we analyzed the three time blocks separately.
  
    We considered only components found in at least four epochs. This results in only one derived velocity for the eastern jet during block~2: ej~10. Together with component wj15 its maximum separation from the center is $< 1$\,mas and shows a low velocity compared to the other tracked components. It is difficult to decouple their properties from the core region. As a result, both components will not be considered in our analysis. Therefore, we do not derive a mean speed for the eastern jet in block~2. 

    The overall mean, maximum, and minimum values for each jet and the mean values for each time block and both jets are listed in Tab. \ref{tab:timeevo_fits}. The uncertainties represent the standard deviation from the mean, assuming an uncertainty on the position of the components as defined in Section~\ref{sec:Observation_Reduction}. We derived $\beta_\mathrm{wj}=0.343\pm0.037$ and $\beta_\mathrm{ej}=0.529\pm0.038$ for the western and eastern jet, respectively.
   
    \begin{table}[!ht]
     \setlength{\extrarowheight}{1.5pt}
     \setlength{\tabcolsep}{3pt}
     \small
     \centering
     \caption{Averaged speeds for the western jet (top) and eastern jet (bottom) for each observational time block and mean, minimum and maximum for all observations for each jet.}
     \label{tab:timeevo_fits}
     \begin{tabular}{lcc}\hline\hline
      & \multicolumn{2}{c}{Western Jet}\\
      & $v_\mathrm{mean}\,[\mathrm{mas\,yr^{-1}}]$& $\beta_\mathrm{mean}\, [c]$ \\\hline
      Block 1 & 1.361 $\pm$ 0.230 & 0.412 $\pm$ 0.069  \\
      Block 2 & 0.882 $\pm$ 0.074 & 0.267 $\pm$ 0.023  \\
      Block 3 & 0.981 $\pm$ 0.093 & 0.297 $\pm$ 0.028  \\
      Mean & 1.134 $\pm$ 0.122 & 0.343 $\pm$ 0.037  \\
      Min & 0.753 $\pm$  0.048 & 0.228 $\pm$ 0.016 \\
      Max & 2.152 $\pm$  0.138 & 0.651 $\pm$ 0.042 \\
      \hline
     \end{tabular}\vspace{0.2cm}
   
     \begin{tabular}{lcc}\hline\hline
      & \multicolumn{2}{c}{Eastern Jet}\\
      &$v_\mathrm{mean}\,[\mathrm{mas\,yr^{-1}}]$& $\beta_\mathrm{mean}\, [c]$ \\\hline
      Block 1 & 2.025 $\pm$ 0.060 & 0.613 $\pm$ 0.018 \\
      Block 2 & -- & -- \\
      Block 3 & 1.336 $\pm$ 0.125 & 0.404 $\pm$ 0.038 \\
      Mean & 1.749 $\pm$ 0.126 & 0.529 $\pm$ 0.038 \\
      Min & 1.181 $\pm$ 0.098& 0.357 $\pm$ 0.030 \\
      Max & 2.156 $\pm$ 0.055 & 0.652 $\pm$ 0.017 \\\hline
     \end{tabular}
    \end{table}
  
    \begin{table}[!ht]
     \setlength{\extrarowheight}{1.5pt}
     \setlength{\tabcolsep}{3pt}
     \small
     \centering
     \caption{Calculated velocities (v) and ejection times (ejt) for the individual components including including errors.}
     \label{tab:speeds}
     \begin{tabular}{lccc}\hline\hline
	     ID   & v[mas/yr]    &   $\boldsymbol \beta_\mathrm{app}$[c]  &    ejt[yr] \\\hline
      ej1 &   2.116 $\pm$ 0.105 &     0.640 $\pm$ 0.032 &     2004.945 $\pm$ 0.043\\
      ej2 &   2.156 $\pm$ 0.055 &     0.652 $\pm$ 0.017 &     2005.168 $\pm$ 0.016\\
      ej3 &   2.034 $\pm$ 0.140 &     0.615 $\pm$ 0.042 &     2005.513 $\pm$ 0.040\\
      ej4 &   1.906 $\pm$ 0.125 &     0.577 $\pm$ 0.038 &     2005.543 $\pm$ 0.029\\
      ej5 &   2.145 $\pm$ 0.141 &     0.649 $\pm$ 0.042 &     2005.841 $\pm$ 0.022\\
      ej6 &   1.790 $\pm$ 0.253 &     0.542 $\pm$ 0.076 &     2005.798 $\pm$ 0.056\\
      ej7 &   -- &     -- &     --\\
      ej8 &   -- &     -- &     --\\
      ej9 &   -- &     -- &     --\\
      ej10 &  0.451 $\pm$ 0.101 &     0.137 $\pm$ 0.031 &     2006.389 $\pm$ 0.263\\
      ej11 &  1.710 $\pm$ 0.095 &     0.517 $\pm$ 0.029 &     2007.695 $\pm$ 0.043\\
      ej12 &  1.245 $\pm$ 0.128 &     0.377 $\pm$ 0.039 &     2008.162 $\pm$ 0.055\\
      ej13 &  1.181 $\pm$ 0.098 &     0.357 $\pm$ 0.030 &     2008.560 $\pm$ 0.050\\
      ej14 &  1.208 $\pm$ 0.131 &     0.365 $\pm$ 0.040 &     2008.844 $\pm$ 0.049\\
      ej15 &  -- &     -- &     --\\
      wj1 &   $-$2.152 $\pm$ 0.138 &    0.651 $\pm$ 0.042 &     2004.322 $\pm$ 0.072\\
      wj2 &   $-$1.600 $\pm$ 0.032 &    0.484 $\pm$ 0.010 &     2004.611 $\pm$ 0.024\\
      wj3 &   $-$0.753 $\pm$ 0.048 &    0.228 $\pm$ 0.015 &     2003.446 $\pm$ 0.157\\
      wj4 &   $-$0.760 $\pm$ 0.077 &    0.230 $\pm$ 0.023 &     2003.946 $\pm$ 0.167\\
      wj5 &   $-$1.727 $\pm$ 0.096 &    0.523 $\pm$ 0.029 &     2005.571 $\pm$ 0.032\\
      wj6 &   $-$1.177 $\pm$ 0.099 &    0.356 $\pm$ 0.030 &     2005.698 $\pm$ 0.042\\
      wj7 &   -- &     -- &     --\\
      wj8 &   $-$1.031 $\pm$ 0.074 &    0.312 $\pm$ 0.022 &     2005.906 $\pm$ 0.104\\
      wj9 &   $-$0.814 $\pm$ 0.190 &    0.246 $\pm$ 0.057 &     2006.344 $\pm$ 0.260\\
      wj10 &  $-$0.802 $\pm$ 0.130 &    0.243 $\pm$ 0.039 &     2007.190 $\pm$ 0.065\\
      wj11 &  $-$0.784 $\pm$ 0.074 &    0.237 $\pm$ 0.022 &     2006.607 $\pm$ 0.204\\
      wj12 &  $-$1.104 $\pm$ 0.042 &    0.334 $\pm$ 0.013 &     2007.794 $\pm$ 0.042\\
      wj13 &  $-$1.173 $\pm$ 0.056 &    0.355 $\pm$ 0.017 &     2008.235 $\pm$ 0.037\\
      wj14 &  $-$0.862 $\pm$ 0.069 &    0.261 $\pm$ 0.021 &     2008.496 $\pm$ 0.052\\
      wj15 &  $-$0.435 $\pm$ 0.017 &    0.132 $\pm$ 0.005 &     2008.573 $\pm$ 0.023\\
      \hline   
     \end{tabular}
    \end{table}
  
 \subsection{Brightness temperature and opening angle}
   \label{sec:brightness_opening}
 
   Following the notation of \cite{Kad04b}, the brightness temperature gradient along the jet axis can be approximated by a power law $T_\mathrm{b}\propto r^s$ with the power-law index $s=d+n+b(1-\alpha_s)$, where $\alpha_S$ is the spectral index and assuming that the magnetic field $B$, the electron density $N$, and the jet diameter $D$ are described by power laws as $B\propto r^b$, $N\propto r^n$  and $D\propto r^d$.

   Orthogonal distance regression fits were applied to the brightness temperature and component size of the Gaussian model components with distance from the core. Given the straightness of the jets and the good agreement of jet width with model components, the size of the model components is assumed to represent the width of the jets. In order to avoid confusion from the core region, all components closer than 0.2\,mas to the core were excluded. There are a few components with unphysical delta-like sizes leading to a very hight brightness temperature, which exceeds the inverse Compton limit. These had been excluded as well. We assigned uncertainties as discussed in Section~\ref{sec:Observation_Reduction}.
   
   \begin{figure*}[!ht]
    \centering
    \begin{minipage}[c]{0.67\textwidth}
     \includegraphics[width=1\linewidth]{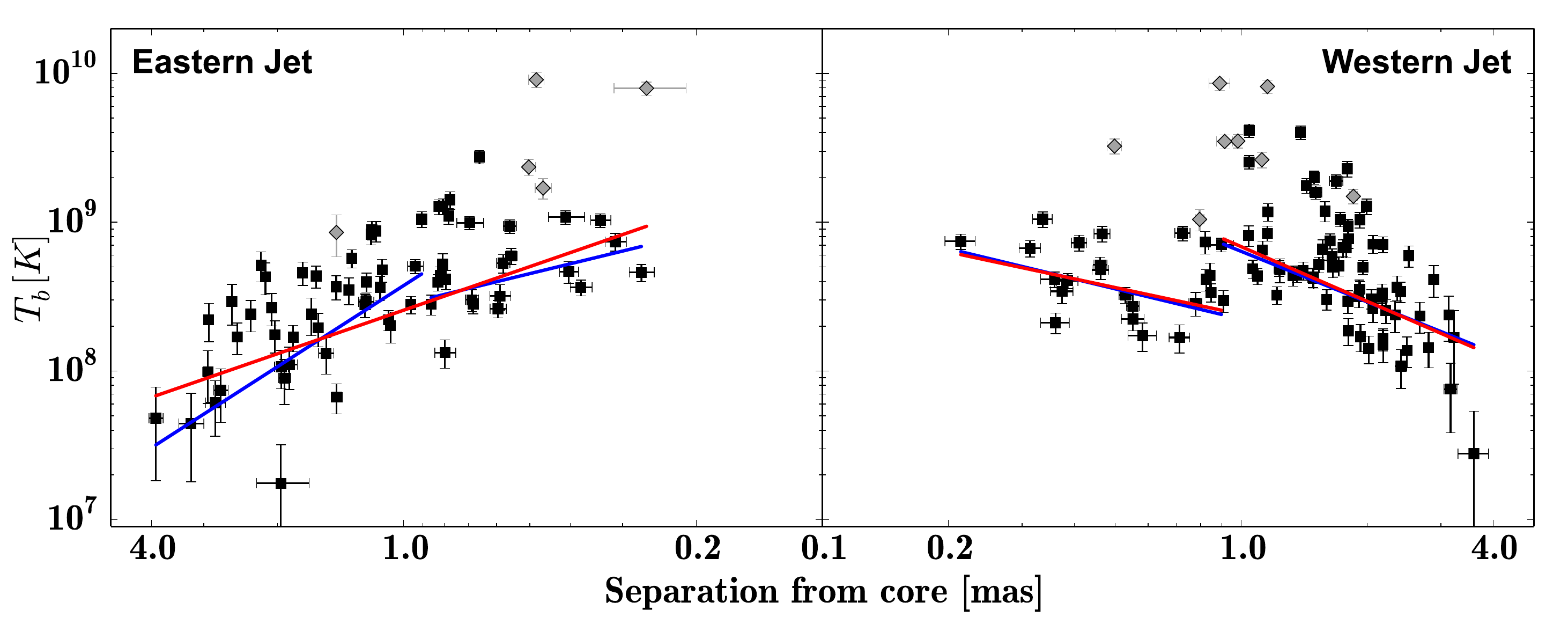}
 
     \includegraphics[width=1\linewidth]{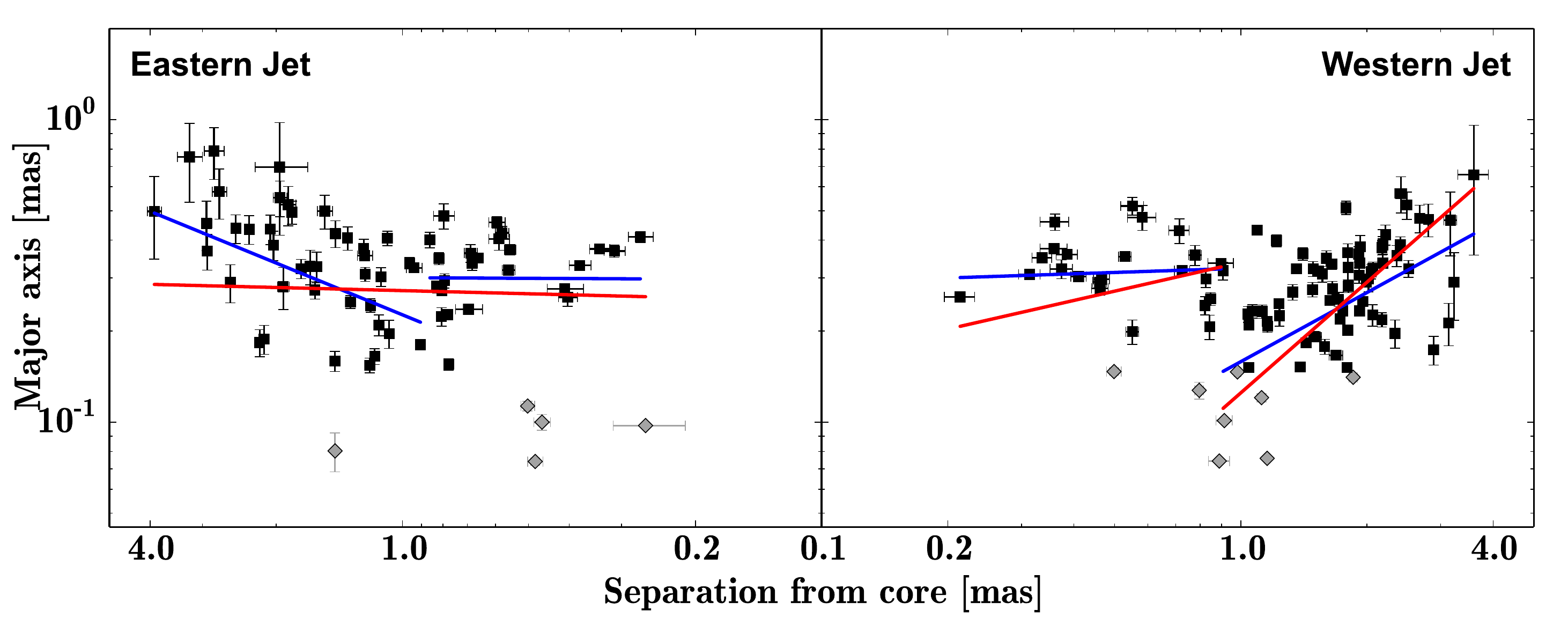}
    \end{minipage}\hspace{0.03\textwidth}
    \begin{minipage}[r]{0.27\textwidth}
     \caption{Brightness temperature (top) and component major axis (bottom) as function of the core distance for both jets. Power laws have been fitted ($T_b\propto r^{s}\,;\;D\propto r^d$) and the indices are printed in Tab.~\ref{tab:s_l}. Two cases have been distinguished: including all data points  (red lines) and excluding the gray diamonds, corresponding to circular Gaussian fits whose major axes are smaller than 0.15\,mas in size (blue lines). These data points do not seem to be a good representation of the width of the jets, but rather maybe due to substructures within the jets.}
     \label{fig:ditb}
    \end{minipage}
   \end{figure*}
 
   As seen in Fig.~\ref{fig:ditb}, the brightness temperatures, and jet diameter do not show a simple power-law like behavior. However, we observe trends in the gradients $s$ and $d$ for the brightness temperature and the opening angle, respectively. When fit with a single power law from 0.2\,mas to 4\,mas we derive the following index values for the eastern jet: $s_\mathrm{ej}=-0.97\pm 0.15$ and $d_\mathrm{ej}=0.03\pm0.09$.
   The western jet shows a break in the $T_\mathrm{b}$ distribution and jet diameter. In the outer part, the slopes are steeper than in the inner part. Applying two power laws (red lines in Fig.~\ref{fig:ditb}) from 0.2\,mas to 0.9\,mas and from 0.9\,mas to 4.0\,mas results in the parameters listed in Tab.~\ref{tab:s_l}, suggesting a conical expansion in the outer part for the western jet.
   
   \begin{table}[!h]
    \setlength{\extrarowheight}{1.5pt}
    \setlength{\tabcolsep}{3pt}
    \small
    \centering
    \caption{Power-law indices for brightness temperature ($T_b\propto r^s$) and jet diameter ($D\propto r^d$) of both jets. Orthogonal distance regression fits had been applied to the different regions as shown in Fig.~\ref{fig:ditb}.}
    \label{tab:s_l}
    \begin{tabular}{lcccc}\hline\hline
    \multicolumn{5}{c}{Fit to all data points}\\\hline
    Region [mas]	&  $s_\mathrm{wj}$ & $s_\mathrm{ej}$ & $d_\mathrm{wj}$ & $d_\mathrm{ej}$\\\hline
    0.2 to 4.0 & -- & $-0.97\pm 0.15$ & -- & $0.03\pm0.09$\\
    0.2 to 0.9 &$-0.60\pm 0.40$& -- & $0.32\pm0.19$ & --\\
    0.9 to 4.0 &$-1.22\pm 0.29$& -- &$1.22\pm0.16$ & --\\\hline
    \multicolumn{5}{c}{Fit to robust data points}\\\hline
    0.2 to 0.9 &$-0.68\pm 0.34$& $-0.68\pm0.37$ & $0.05\pm0.07$ & $0.01\pm0.10$\\
    0.9 to 4.0 &$-1.13\pm 0.27$& $-1.81\pm0.39$ & $0.76\pm0.14$ & $0.57\pm0.20$\\\hline
    \end{tabular}
   \end{table}
   
   There are several components with small sizes that do not seem to be a good representation of the width of the jets, but rather maybe due to local substructures within the jets (see Fig.~\ref{fig:TimeEvo1} and \ref{fig:TimeEvo2}). We excluded all model-fit components with sizes below 0.15\,mas (grey diamonds in Fig.~\ref{fig:ditb}) and fit power laws to both regions (0.2 to 0.9\,mas and 0.9 to 4\,mas) for both jets (blue lines in Fig.~\ref{fig:ditb}). The power law indices are listed in the bottom part of Tab.~\ref{tab:s_l}.

\section{Discussion}
 \label{sec:Discussion}
 
 \subsection{Torus}
  \label{sec:Torus}

  The stacked spectral-index map (see Section~\ref{sec:Spectral_analysis} and Fig.~\ref{fig:sm_stacked}) reveals an absorbing structure near the base of the two jets. The absorber covers a region of about $ 0.4\,$mas (0.04\,pc) to the east and $0.8\,$mas (0.08\,pc) to the west from the 43\,GHz map peak. The spectral index $\alpha$ in this region exceeds the theoretical upper limit for synchrotron self-absorption of 2.5. The 43\,GHz core is very close to the peak of the absorbing structure and therefore we do not expect a significant offset from the dynamic center.
  Furthermore, both jets show a relatively steep spectrum, with a mean spectral index around $-1$ which further decreases down to a minimum of $\sim -2$. This is in agreement with the values reported by \cite{Kad04b} of around $-1$ in both jets, which suggests that the outer jets are optically thin.
  By using the stacked map it is possible to produce images at both frequencies that show no gap in the emission. This allows us to give a more continuous measurement of the geometry of the absorbing medium than possible in previous works \cite[see e.g.,][]{Kellermann:1999,Kameno2001,Vermeulen2003,Kad04b}. 
  
 \citep{Kameno2001} derived the distribution of the free-free absorption (FFA) opacity due to the torus and found a peak opacity at 1\,GHz of $\tau_\mathrm{f,\,1\,GHz}=300$ in the nucleus. The FFA opacity coefficients derived by \cite{sawada-satoh2008} are consistent with this, but show a higher mean optical depth in the center of $\tau_\mathrm{f,\,1\,GHz}=1000$. Given a frequency dependence of the optical depth as $\tau_\mathrm{f}\propto\nu^{-2.1}$ the free-free opacity at 43\,GHz can be estimated to $\tau_\mathrm{f,\,43GHz}=0.11$ for assuming $\tau_\mathrm{f,\,1\,GHz}=300$ and to $\tau_\mathrm{f,\,43GHz}=0.37$ for assuming $\tau_\mathrm{f,\,1\,GHz}=1000$. The observed flux density at $43$\,GHz ($S_\mathrm{abs}(43\,\mathrm{GHz})$) is typically around $300\,$mJy for the core region. It is connected to the intrinsic flux density ($S_\mathrm{int}(\nu)$) and $\tau_\mathrm{f}$ by $S_\mathrm{abs}(\nu)=S_\mathrm{int}(\nu)\,\mathrm{e}^{-\tau_\mathrm{f}}$. For $S_\mathrm{abs}(43\,\mathrm{GHz})=300\,$mJy we get $S_\mathrm{int}(43\,\mathrm{GHz}) =335\,$mJy and $S_\mathrm{int}(43\,\mathrm{GHz}) =434\,$mJy. 
Based on these assumptions we cannot assume this region to be optically thin. However, there are very few epochs in which jet components are close enough to the central core to be affected by absorption (for example BR130F or BR130E). These epochs reveal strongly correlated Gaussian model parameters in the innermost jet regions including the core. Therefore, these observations are already not optimal to analyse the innermost region. For all other observations the central component will still be the brightest one when accounting for absorption and therefore still be the most likely position of the core.

  The location of the cut in the power laws in the western jet, as seen in Fig.~\ref{fig:ditb}, is consistent with the outer edge of the absorbing torus. 
  If we assume still some absorption at 43\,GHz, this would result in a reduction of the flux density of the western jet in this region, which would explain satisfactorily the brightness temperature distribution \cite[cf.][]{Kad04a}. If the drop in the brightness temperature distribution is only due to the absorption, interpolating the outer gradient inwards should give us the \textit{real} brightness temperatures close to the central region. This results in higher values for the western jet.
  In addition, the asymmetry observed from 2007 on will be even more pronounced.

  In comparison, \cite{Kad04b} derived power law index values $s$ between $-3.8\pm0.3$ and $-4.1\pm0.8$ for the eastern jet at frequencies of 5, 8.4, 22, and 43\,GHz. 
  The brightness temperature distribution of the western jet could not be approximated by a simple power law. 
  The fits were consistent for all frequencies but were performed only for components farther away than 2.5\,mas from the center. 
  \cite{Kad04b} analyzed only one epoch per frequency, whereas our analysis includes 26 epochs at 43\,GHz, resulting in a more robust statistics. Adding the derived values by \cite{Kad04b} to our analysis, the absolute value of the $T_\mathrm{b}$ power-law index increases with increasing distance from the core. 

In the numerical simulations of \cite{Fro18} initially symmetric jets are assumed. However, due to the delicate interplay between absorption and orientation of the jet-torus system, and the properties of the observing array, asymmetries between the jet and counter jet appear.

In the case of over-pressured jets embedded in a decreasing pressure ambient medium several recollimation shocks can be formed. The observed signature of a recollimation shock can be described by a stationary local flux density maximum (in contrast to travelling shocks where the local increase in flux density is advected with the plasma). The local increase in pressure and density at the recollimation shocks decreases with distance from the jet nozzle and thus the observed emission signatures of standing recollimation shocks diminish with distance down the jet. Therefore, the strongest recollimation shocks in each jet ( which would appear as stationary flux density maxima) should be located to the jet nozzle.

However, depending on the torus properties, the observing frequency and the orientation of the jet-torus system these features can be hidden behind the obscuring torus. The obscuring torus becomes optically thinner at higher frequencies offering us a glimpse into the innermost regions of the central engine. As a consequence the local flux density maxima associated with the recollimation shocks may become visible \cite[see Fig.14 in][]{Fro18}. If we assume \object{NGC\,1052} has a viewing angle $\vartheta\neq90$ we expect the larger absorption along the line of sight to result in less flux in one of the jets (i.e. in the counter jet).
Also large spectral index values ($\sim 3$) can be obtained between the jet and counter jet due to the contribution of the free-free absorption in the torus \cite[see Fig. 16 in][]{Fro18}. 

To test whether the jets are initially symmetric or asymmetric detailed simulations and modelling tailored to \object{NGC\,1052} have to be performed.
   
 \subsection{Jet properties}
  \label{sec:speed_viewing}
  \subsubsection{Morphology}

Comparing the first epoch at 43\,GHz in 2005 and the 86\,GHz map from October 2004 \cite{Baczko16}, both maps reveal one central bright feature and two fainter, symmetrically evolving jets. We attempted component identification between both frequency maps. However, there are several difficulties that prevent us from obtaining a robust result: 1) There is a gap of 5 month between the 86\,GHz observation and the first 43\,GHz observation. Based on our study, the morphology of the source might change significantly during this time; 2) Extrapolating the western jet component locations to the time of the 86\,GHz observation, there are several paths which are crossing. Therefore, it is not clear where the components had been located in October 2004; 3) Extrapolating the components in epoch BR099A backwards gives only information about the inner 1\,mas of the eastern jet, which is not much to compare with at 86\,GHz; 4) The quality of the 86\,GHz data is not good enough to make a reliable association. For this, one should consider a model fit degrading the weights of the long-baseline data (tapering), since we do not have enough signal-to-noise ratio in the full-resolution image at 3.5\,mm.

    As seen in Figures \ref{fig:TimeEvo1} and \ref{fig:TimeEvo2}, there are remarkable differences between the first observational block and the following two. Therefore, stacked maps for each block have been produced (see Fig.~\ref{fig:RidgeLine2}). 
    In block~1 \object{NGC\,1052} shows a symmetric morphology, whereas in block~2 the jets become asymmetric and the source is more compact. The extent of the western jet is about 1\,mas smaller than in block~1. The eastern jet becomes less prominent in blocks~2 and 3 (see Fig.~\ref{fig:RidgeLine2}). It shows only diffuse emission beyond 2\,mas.
    \newline
    The fits to the jet diameter indicate strict deviations from a conical expansion in the inner 1\,mas (corresponding to 0.1\,pc) of both jets. At the point of the change in the power law index, there is a break in the distribution, resulting in smaller major axes at around 1\,mas, indicating a possible re-collimation of the jets.

  \subsubsection{Speed}
    There is a clear trend toward higher velocities in the eastern jet in comparison to the western jet (see Tab.~\ref{tab:speeds}). For time block\,1 the basic kinematic model was previously presented in \cite{Ba16b}. The western jet in block\,1 shows a wide spread of velocities. In contrast to that the component velocities in the eastern jet are quite similar, with deviations of less than $0.1\,c$ from the mean value. For block\,2 and block\,3 the western jet is more stable, revealing comparable speeds for all tracked components. However, the overall evolution in both jets show that, besides the clear trend towards higher velocities in the eastern jet, there are significant deviations from the mean speed. 
  
\begin{figure}
     \centering
     \includegraphics[width=0.9\linewidth]{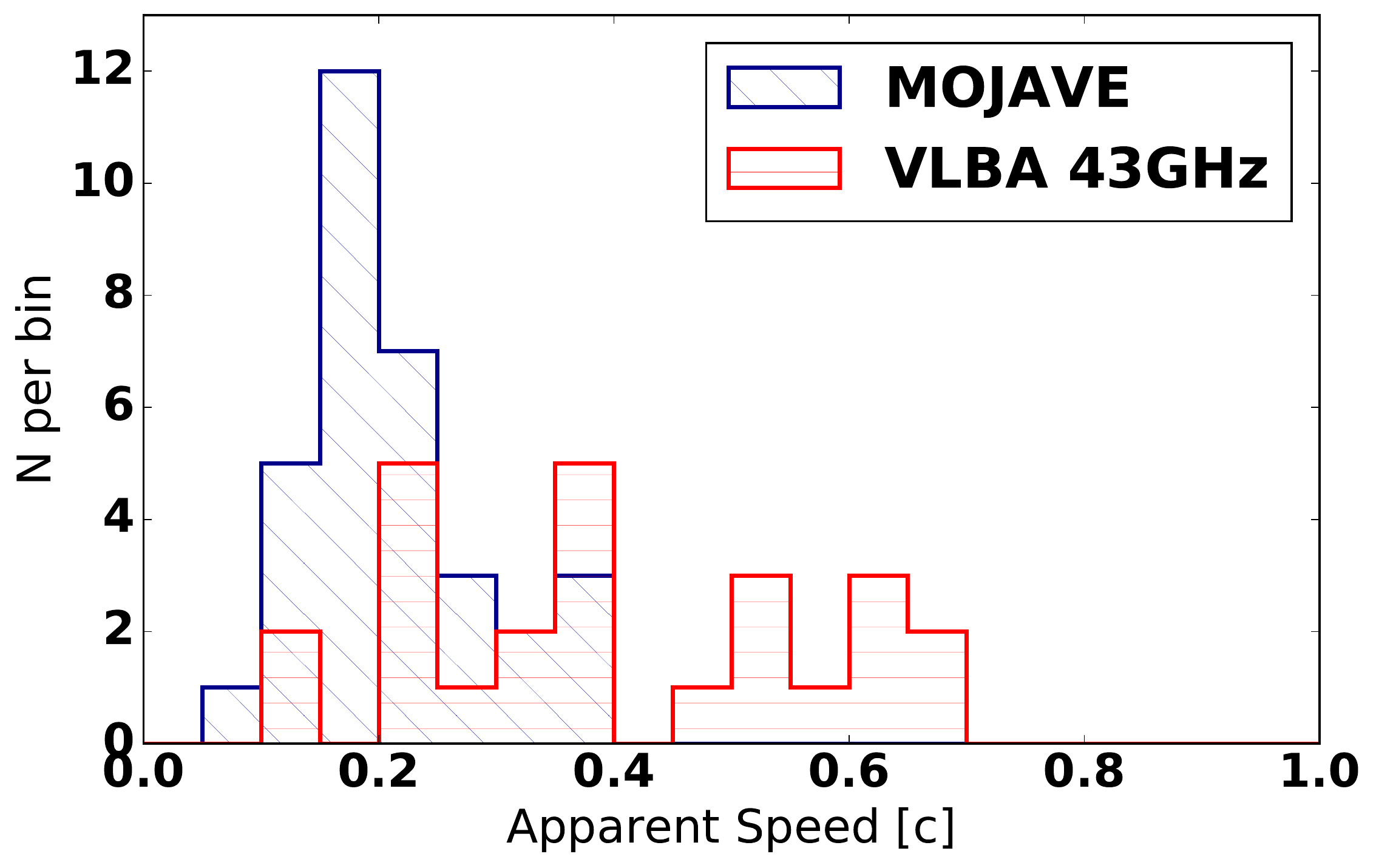}
     \caption{Histogram of the apparent velocity of the jet components for this work at 43\,GHz (red), and \citet{Lis13} (blue, MOJAVE) at 15\,GHz. The bin size is $0.05\,$c.}
     \label{fig:hist_vel}
    \end{figure}

    We compare the measured apparent velocities at 15\,GHz from \citet{Lis13} with the values at 43\,GHz in Fig.~\ref{fig:hist_vel}. 
    The components at 43\,GHz have significantly different values than at 15\,GHz with a tendency of higher velocities at higher frequency. 
    As discussed in the previous sections, a circumnuclear torus covers large parts of the western jet, thus free-free absorption results in an obscuration of the inner part at lower frequencies \citep{Kameno2001,Vermeulen2003,Kad04b}. Its impact at 43\,GHz is already small enough so that we are able to peer through this absorbing structure. If the jet has transversal velocity structure, it is likely, that we are observing a faster moving layer of the jets at 43\,GHz as compared to lower frequencies. The slower moving layer of the jet, as observable at 15\,GHz, may very well be invisible at 43\,GHz due to less energetic particles which are not emitting at 43\,GHz.
    
  \subsubsection{Viewing angle}
  \label{sec:viewing_angle}
    An estimate on the angle of the jets to the line of sight $\theta_\mathrm{LOS}$ can be derived from their flux-density ratio. If we assume that both jets are intrinsically symmetric, and evolve with an angle of $180\,$deg between the two jet axes:

    \begin{equation}
     R=\frac{S_\mathrm{approaching}}{S_\mathrm{receding}} = \left(\frac{1+\beta\cos\theta_\mathrm{LOS}}{1-\beta\cos\theta_\mathrm{LOS}}\right)^{2-\alpha}\quad.
     \label{eq:LOS}
    \end{equation}
    Here $\beta$ is the intrinsic jet speed and $\alpha$ the spectral index. According to Section~\ref{sec:Torus} we adopt a value of $\alpha=-1$, consistent with our 22-43\,GHz observations.

    In addition, the angle of both jets to the line of sight can be derived with knowledge of their intrinsic ($\beta$) and apparent ($\beta_\mathrm{app}$) speeds for approaching and receding jets:
    \begin{equation}
	     \beta_{\rm app,approaching/receding} = \frac{\beta \sin \theta_{\rm LOS}}{1 \mp \beta \cos \theta_{\rm LOS}}\quad .
     \label{eq:Beta}
    \end{equation}
    
    Both equations can be combined to derive the allowed parameter space for the intrinsic velocity $\beta$ and the angle of the jets to the line of sight $\theta_\mathrm{LOS}$ by using the derived values for the velocity and flux density ratios.
    
    As discussed in the previous sections, there are differences in velocity between both jets and the different time blocks. In addition we find the western jet to be consistently brighter than the eastern jet at 43\,GHz. This result is contradictory to earlier studies, that showed the eastern jet to be the brighter one \citep{Kellermann:1999,Vermeulen2003,Kad04a}. Our analysis reveals a change in the jet-to-counter-jet ratio over the span of four years from around $R=1.1$ up to $R=2.5$ (see Fig.~\ref{fig:lightcurve}c). For the first observational block, the ratio remains at a similar level of around $R=1.1$, in this period the morphology of the jets appears to be symmetric. In the second observational block (2007) the jets become increasingly asymmetric with a mean ratio of about $R=1.9$.
    
\begin{figure}[!ht]
      \centering
      \includegraphics[width=0.95\linewidth]{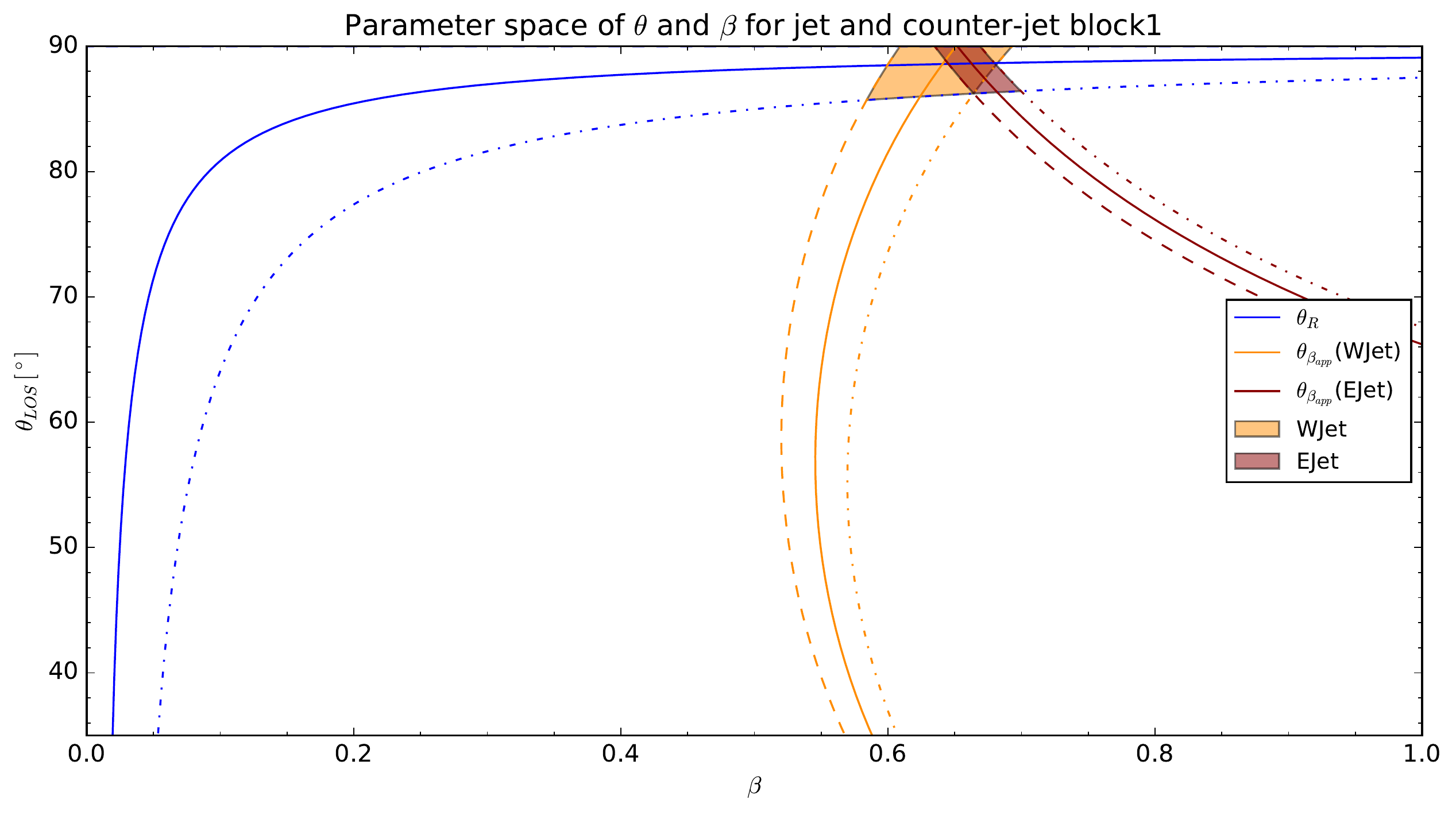}     	 
      \includegraphics[width=0.95\linewidth]{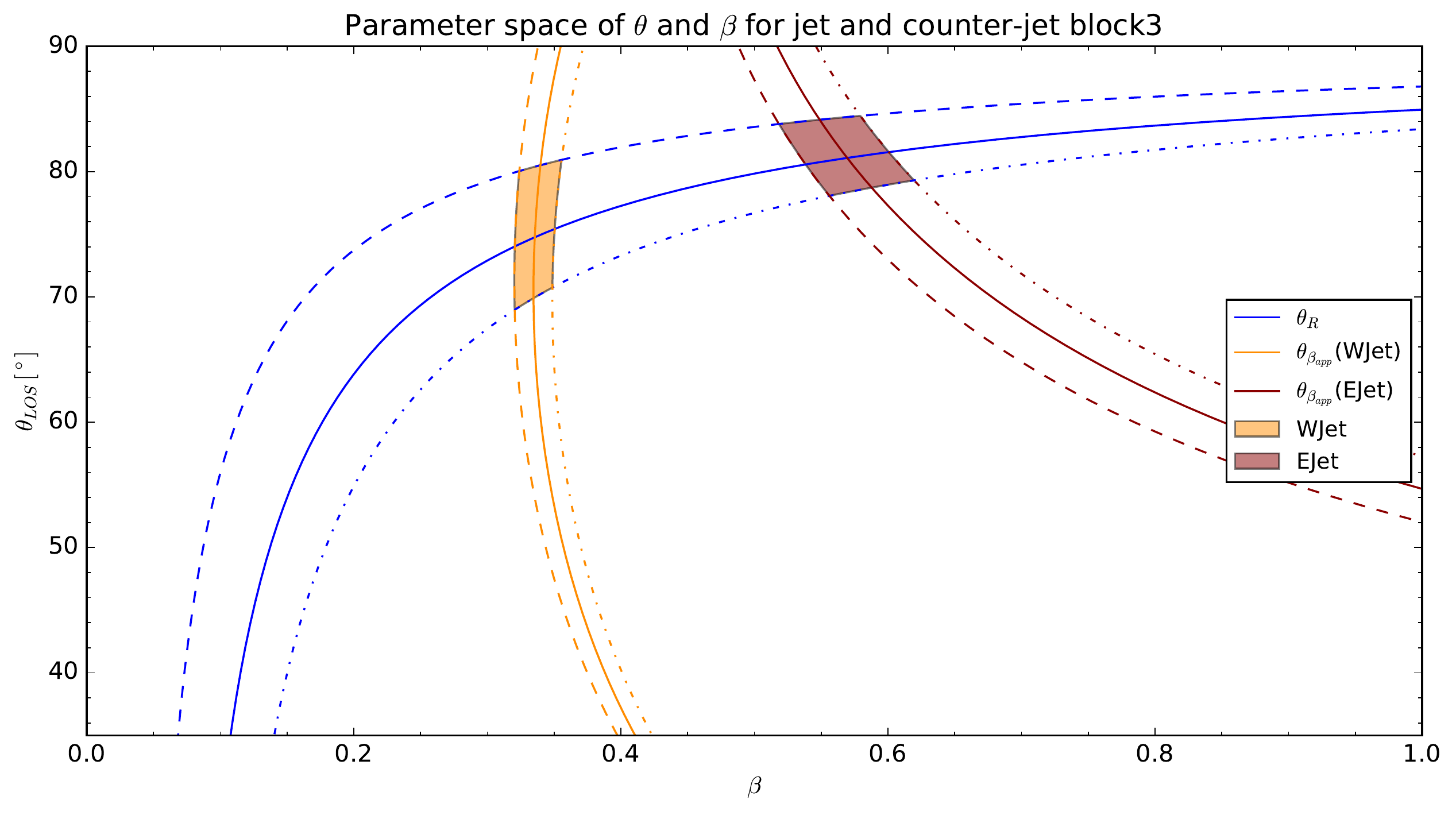} 
      \caption{ The angle of the jets to the line of sight $\theta_\mathrm{LOS}$ as a function of the intrinsic jet velocity $\beta$, constrained by the western-to-eastern jet ratio R (blue) and the apparent maximum velocities for eastern jet(red) and western jet (orange) for observational Block~1 and 3. The allowed parameter space for $\theta_\mathrm{LOS}$ and $\beta$ for eastern and western jet is highlighted by the red/orange-shaded region, respectively. 
The dashed and dashed-dotted lines mark the uncertainty range.
We do not find a common parameter space for both jets and both blocks. 
The assumption of intrinsic asymmetry is therefore contradictory to our results.} 
      \label{fig:beta_theta_block1_3}
     \end{figure}

\begin{figure*}
     \centering

     \includegraphics[width=0.33\linewidth]{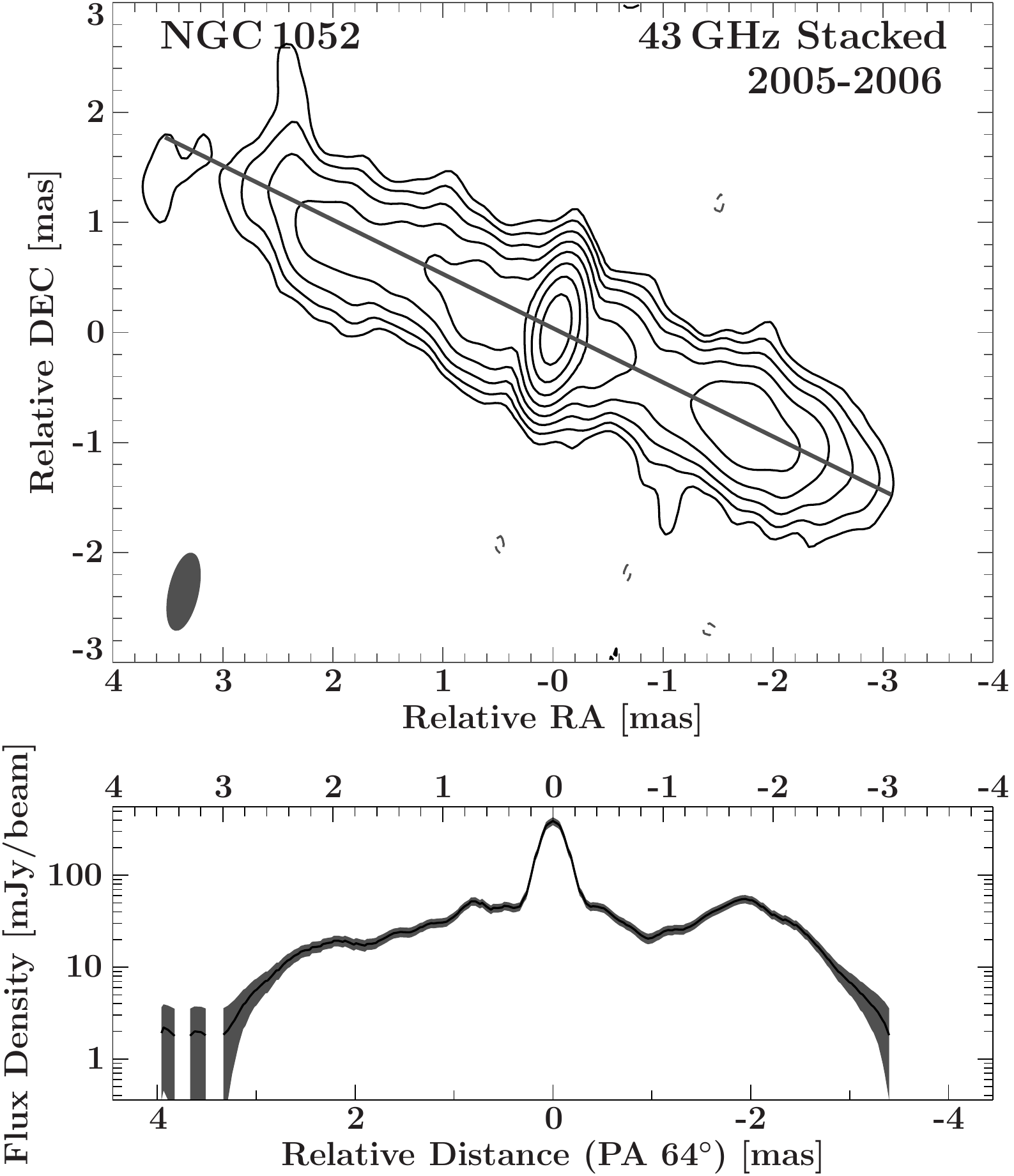}
     \includegraphics[width=0.33\linewidth]{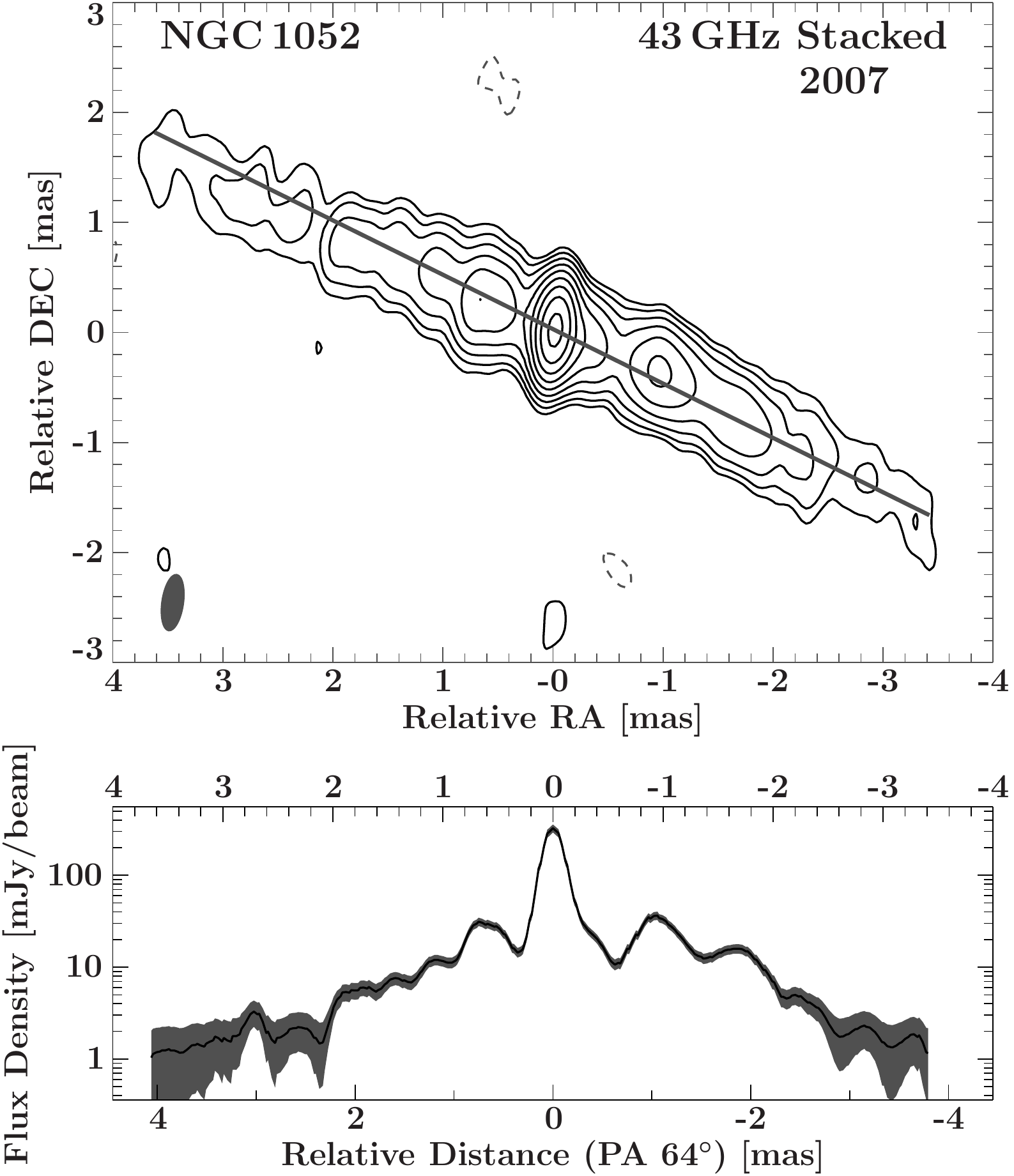}
     \includegraphics[width=0.33\linewidth]{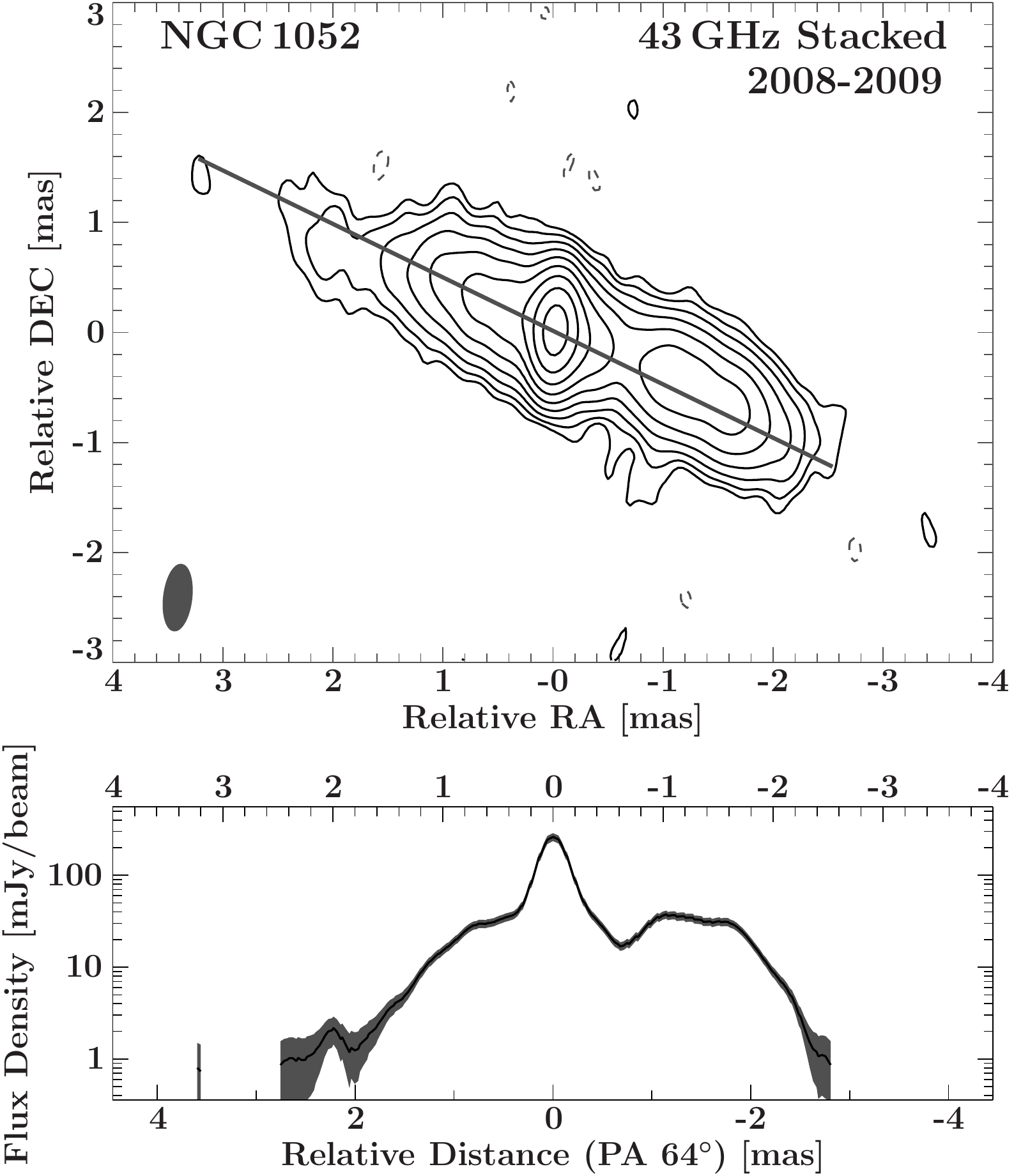}

     \caption{Stacked images (top) with the flux densities along the position angle of the jet (\textit{bottom panels}) for the observations from 2005 - 2006 (\textit{left panel}), 2007 (\textit{middle panel}) and 2008--2009 (\textit{right panel}). The cut along the position angle is indicated in the \textit{upper panels} as a line along the jet axis. The common beam for the stacked maps is shown in the lower left corner of the \textit{top panels}. For the stacked images the maps have been summed up and divided by the number of images. The contours start at 3 times the noise level and increase logarithmically by factors of 2. 
     }
     \label{fig:RidgeLine2}
    \end{figure*}

    Based on our results there is evidence that the jets are not symmetric. However, equations~\ref{eq:LOS} and \ref{eq:Beta} can be used to test this assumption. If symmetry holds, we should be able to obtain a common region in the parameter space for $\theta_\mathrm{LOS}$ and $\beta$ that hold for all observations. The eastern jet in observational block~2 is quite short and reveals only very few tracked components, therefore we compare blocks 1 and 3. In Eq.~\ref{eq:Beta} we used the maximum speeds found during the appropriate blocks. 
    Based on the herein discussed 43GHz observations and the morphology outside the central region, we assumed the western jet to be the approaching jet and the eastern jet to be the receding jet. As shown in Fig.~\ref{fig:beta_theta_block1_3} we find a common region in the parameter space for $\theta_\mathrm{LOS}$ and $\beta$ in Block~1. In Block~3 the velocities depart from each other and cannot be compatible, so it is not possible to combine eastern and western jet assuming intrinsic symmetry.

The assumption that both jets are intrinsically identical in terms of velocity, viewing angle and an overall symmetric evolution was implicit in our
study of the parameter space. However, there is no combination of $\beta$ and $\theta_\mathrm{LOS}$ that is consistent for both blocks and jets. This contradicts the assumption of symmetry, and therefore, at least one equality constraint must be lifted.

   \subsection{Source geometry}
    \label{sec:geometry}
     The asymmetries we observe may originate from asymmetries either in the physical nature of the jets themselves or the environment into which they propagate. If we assume the drop in $T_\mathrm{b}$ in the inner part of the western jet is due to the absorbing structure, the interpolation of the outer gradient results in higher brightness temperatures close to the central engine compared to the eastern jet. This is most likely explained by a higher brightness of the western jet, thus requiring a larger jet internal energy and/or magnetic flux. For example, this can be achieved by a more energetic ejection producing larger particle internal energy and/or magnetic flux. This would favor an internal asymmetry in the jets.
     On the other hand, even if we assume that the same energy flux is injected at the nozzle of both jets, there can also be disturbances further downstream, resulting in different energy conversion and leading to a possible higher emissivity of the western jet. Given the large scales on which the jets are interacting with the interstellar medium it is most likely that this environment is not identical for both jets. In particular, stronger dissipation or increased mass-loading by, e.g., stellar winds, can lead to a brighter and slower flow \citep[see, e.g.][]{bow96,pe14}.

    Combining the results from the kinematic analysis at 43\,GHz with previous studies at lower frequencies \citep[see, e.g.][]{Vermeulen2003} further constrains on the jet properties and formation  can be made. The particles emitting at 43\,GHz are assumed to be accelerated from regions of the accretion disc, that are closer to the SMBH. Finally, we expect to observe an inner layer of the jet. This would lead to a faster moving inner layer and a slower outer layer, which is not emitting at 43\,GHz.

    The physical conditions are getting more extreme when coming closer to the central engine. If the direct surrounding of the central SMBH has a larger impact at 43\,GHz, this would explain the different behavior of the source at this frequency in comparison to lower frequencies. The energy conversion processes are more violent and disturbances can happen on shorter time scales.

   \subsubsection{Precession or asymmetric jet production?}
    As discussed in Sec.~\ref{sec:viewing_angle} our results do not fit into a simple picture in which we assume the same intrinsic conditions in both jets. There are two main possibilities capable of explaining this. (1): jets are not intrinsically identical. (2): if $\beta$ does not change, the flux density ratio only depends on $\theta_\mathrm{LOS}$. This would result in a change of $\theta_\mathrm{LOS}$ over the 4 years of observation, implying a precessing jet.
    
   The galactic X-ray binary \object{SS\,433} is an example of an precessing jet. The jet precession occurs within a cone of half-opening-angle of $19.85^\circ$ awith respect to the angle to our line of sight of $79^\circ$, with a period of 162 days \citep{Mar89}. As a larger scale counterparts to \object{SS\,433}, radio galaxies have been observed to reveal helical motions, too. In particular, a considerable fraction of quasars have S-shaped jets (for example \object{3C273} \citealt{Per17} or \object{3C345} \citealt{mat13}). There are several mechanisms leading to a precessing jet. One often discussed scenario is the Bardeen-Petterson \citep{Bar75} effect, which is due to a misalignment of the angular momentum of the accretion disc and the rotating central black hole. Based on this approach jet precession has been successfully described in several blazars \citep[see, e.g.][]{Liu02,Cap04}. Typically very long precession periods are observed.
   
    There is evidence for a recent merger event in \object{NGC\,1052} about $1$-$2$ Gyrs ago \citep{Gor86,Pie05}. Therefore, another likely explanation is a binary black hole, where the companion exerts a torque and therefore results in a precession of the accretion disk, resulting in shorter periods of precession.

There are many studies concerning jet precession (see for example \cite{Moh16}, \cite{Mac16}, \cite{Roz16}, \cite{Qia17} and references therein). Examples are \object{1308+226} with an orbital period of $\simeq 8.5\,$yr in case of a binary black hole model \citep{Bri17}, \object{2MASXJ12032061+1319316} with a precession period of $0.95\times 10^5\,$yr \citep{Rub17} and \object{M\,81} with a period of $7.27\,$yr \citep{Mar11}. 

If the jets are initially asymmetric the question arises which mechanism is able to produce this behaviour. One possible physical scenario is connected to the magnetic field structure in the surrounding disk which is accreted by the black hole. Assuming there are several small scale magnetic field loops with alternating polarity \citep{Par15,Con18}.

For example if we assume a magnetic field configuration in the disk consisting of four magnetic field loops (two loops in the northern hemisphere of the disk and two in the southern hemisphere with alternating polarity). During the accretion of the innermost loops, the magnetic field is opening up due to the differential rotation and leads to a poloidal magnetic field with different polarity in the northern and southern hemisphere \citep[see e.g. Fig.~1 in][]{Par15}.

If we further assume that the polarity in the northern jet is different to the accreted polarity in the northern hemisphere, the ongoing accretion pushes magnetic field lines of different polarity closer and closer together. This leads to a reconnection event and disrupts or turns off the northern jet for some time until it is established again. On the other hand there is no difference in the polarity in the southern hemisphere, i.e. the southern jet and the accreted loop have the same polarity. Therefore no reconnection event occurs which disturbs the jet. The situation is flipped in the accretion of the next magnetic loop pair, i.e. the northern jet is unperturbed while the southern one is disrupted. Within such a scenario asymmetric jets with varying outflow can be produced.
 
A detailed study of the magnetic field in the region of the central engine in \object{NGC\,1052} could shed light on whether this could be an explanation for the observed asymmetry. Until now, it was only possible to derive estimates on the mean magnetic field in the central region. Our recent study of \object{NGC\,1052} observed with the GMVA at 86\,GHz \citep{Baczko16} sets boundaries to the magnetic field at $1\,R_\mathrm{S}$ as $200\,\mathrm{Gauss}<B<8\times 10^4\,\mathrm{Gauss}$. However, we do not have linear polarization data that would allow us to infer the geometry of the magnetic field in NGC\,1052. Future mm-VLBI observations may be able to trace changes in the magnetic field of \object{NGC\,1052}. 
    
    Even without detailed knowledge of the magnetic field geometry in NGC\,1052 our results favor an asymmetric jet production. Subtracting the influence of the absorbing torus we expect higher flux density values for the western jet close to the central engine as compared to the eastern one. On the other hand there is a clear trend towards higher velocities in the eastern jet. These observations fit very well into a picture in which the western jet carries a larger internal energy and/or magnetic flux than the eastern jet, thus becoming brighter at these scales.

    Simultaneous multi-frequency observations will allow us to study in detail absorption mechanisms, external and internal, along both jets, thus probing the local magnetic field distribution. 

\section{Conclusion}
  \label{sec:Conclusion}
 
  In this work we present a detailed VLBI analysis of \object{NGC\,1052} involving twenty-nine epochs from 2005 to 2009 at 43\,GHz. 
\begin{enumerate}
 \item In early 2007 the morphology of the source changes from symmetric to asymmetric with the western jet becoming brighter. Further the flux density ratio of the western to eastern jets varies between $R=1.1$ and $R=2.5$.
 \item A kinematic analysis of the evolution of the various jet components results in mean apparent jet velocities of $\beta_\mathrm{wj}=0.343\pm0.037$ and $\beta_\mathrm{ej}=0.529\pm0.038$ for the western and eastern jets, respectively. Compared to previous estimates at lower frequencies\citep{Vermeulen2003,Lis13} the apparent jet speeds derived in this work are faster, especially for the western jet. 
 \item Besides the extended jet emission, one central bright emission region is present in all observed epochs. Based on the kinematic analysis this central feature is likely the dynamic center of the system.
 \item Combining the flux density evolution and kinematics, we explored the parameter space of the intrinsic velocity and viewing angle. We did not find consistent parameters for all observations, which points towards asymmetric jet production.
  
 \item We analyzed the spectral properties of the source morphology between 22\,GHz and 43\,GHz. It reveals an optically thick region to the west of the brightest jet feature at 22\,GHz with a diameter of around $1.5$\,mas ($0.15$\,pc). The absorption seems to be caused by a torus around the central engine which is perpendicular to the twin-jet system. Both jets are optically thin with spectral indices $\alpha\leq-1$.
  
 \item We fit power laws to the distribution of brightness temperatures and jet diameter as a function of the separation from the center. 
  The expansion is similar for both jets, leading to a well-collimated structure, that evolves, starting at around 1\,mas, with a power law index of $d=0.65$.
 \item The brightness temperature values in the western jet drop at $1\,$mas separation to the center. This is well be explained by a reduction of the flux density due to absorption in the torus.
 \item We propose a faster inner layer that is fed from a region closer to the central engine, as observed at 43\,GHz, and a slower moving outer layer, with particles coming from further outwards of the accretion disc, as observed at lower frequencies.
 \item These observations fit very well into a picture in which the western jet carries a larger internal energy and/or magnetic flux than the eastern jet, thus becoming brighter at these scales. 
\end{enumerate}

  Further studies at 43\,GHz are needed to test the observed asymmetry on longer time scales. As an observation at 86\,GHz resulted in a similar morphology \citep{Ba16b}, simultaneous observations up to the highest achievable radio frequencies would allow us to perform the core-shift analysis and spectral index measurements up to even higher frequencies. This will result in a more complete picture of the twin-jet system in \object{NGC\,1052} and would give us the chance to test current jet formation models.

\begin{acknowledgements} 
M.P.\ acknowledges partial support from the Spanish MICINN grant AYA-2013-48226-C03-02-P and the Generalitat Valenciana grant PROMETEOII/2014/069.
R.S. gratefully acknowledge support from the European Research Council under the European Union's Seventh Framework Programme (FP/2007-2013)/ERC Advanced Grant RADIOLIFE-320745"
CMF is supported by the ERC Synergy Grant ``BlackHoleCam - Imaging the Event Horizon of Black Holes'' (Grant 610058).
This research has made use of the software package \textsc{ISIS}  \citep{Houck2000} and a collection of \textsc{ISIS} scripts provided by the Dr. Karl Remeis Observatory, Bamberg, Germany at \texttt{http://www.sternwarte.uni-erlangen.de/isis/}, as well as an \textsc{ISIS}-DIFMAP fitting setup by C. Grossberger \citep{Gro14}. This research has made use of \texttt{SciPy} \citep{scipy} and the \texttt{ODR} package, based on \cite{Bo90}.
This research has made use of the NASA/IPAC Extragalactic Database (NED) which is operated by the Jet Propulsion Laboratory, California Institute of Technology, under contract with the National Aeronautics and Space Administration.
This research has made use of data obtained with the Very Long Baseline Array (VLBA). The VLBA is an instrument of the National Radio Astronomy Observatory, a facility of the National Science Foundation operated under cooperative agreement by Associated Universities, Inc.

\end{acknowledgements}
\bibliographystyle{aa}
\bibliography{bibliography}

\begin{appendix}
\section{Additional Figures and Tables}

\begin{figure*}[!h]
  \includegraphics[width=0.45\linewidth]{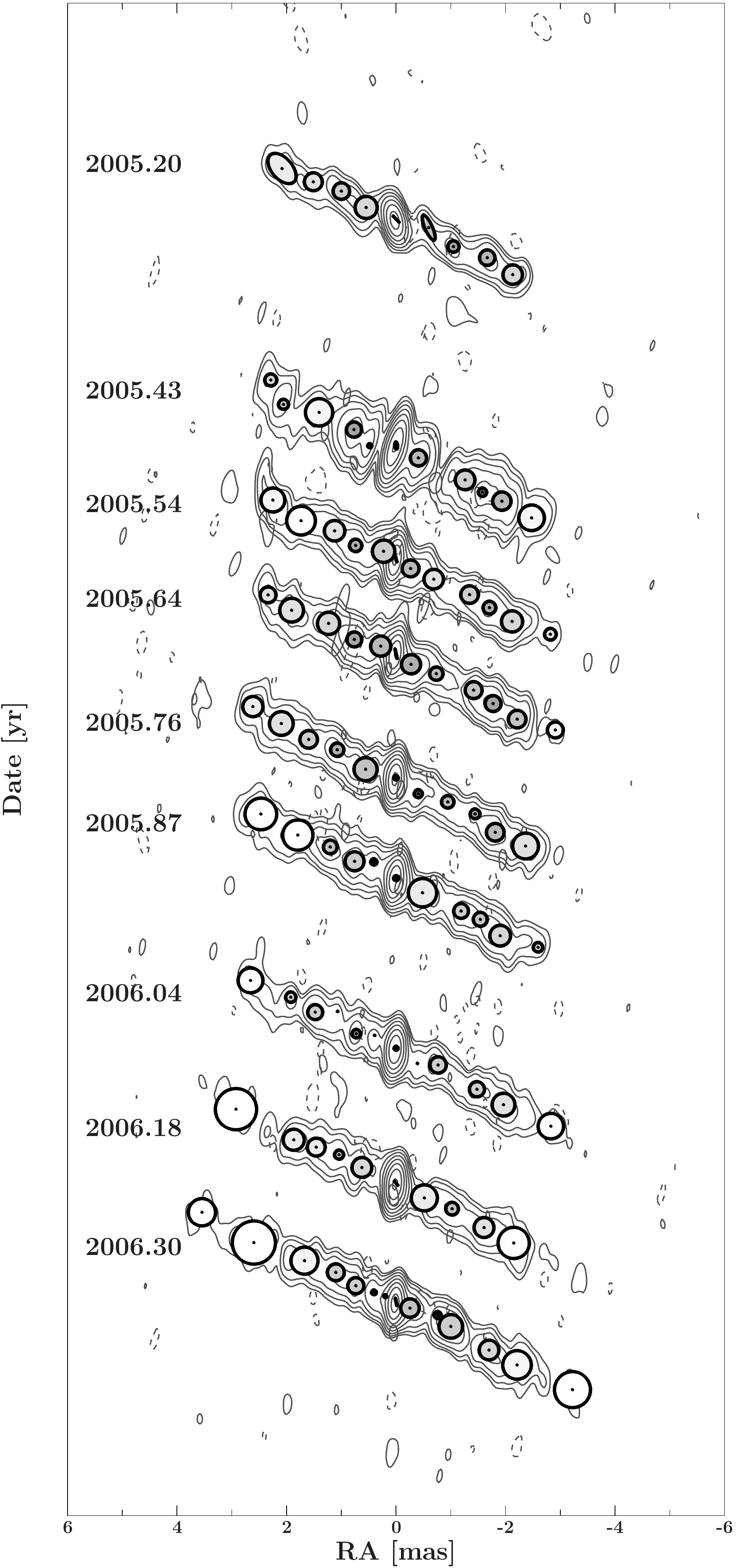}
  \includegraphics[width=0.45\linewidth]{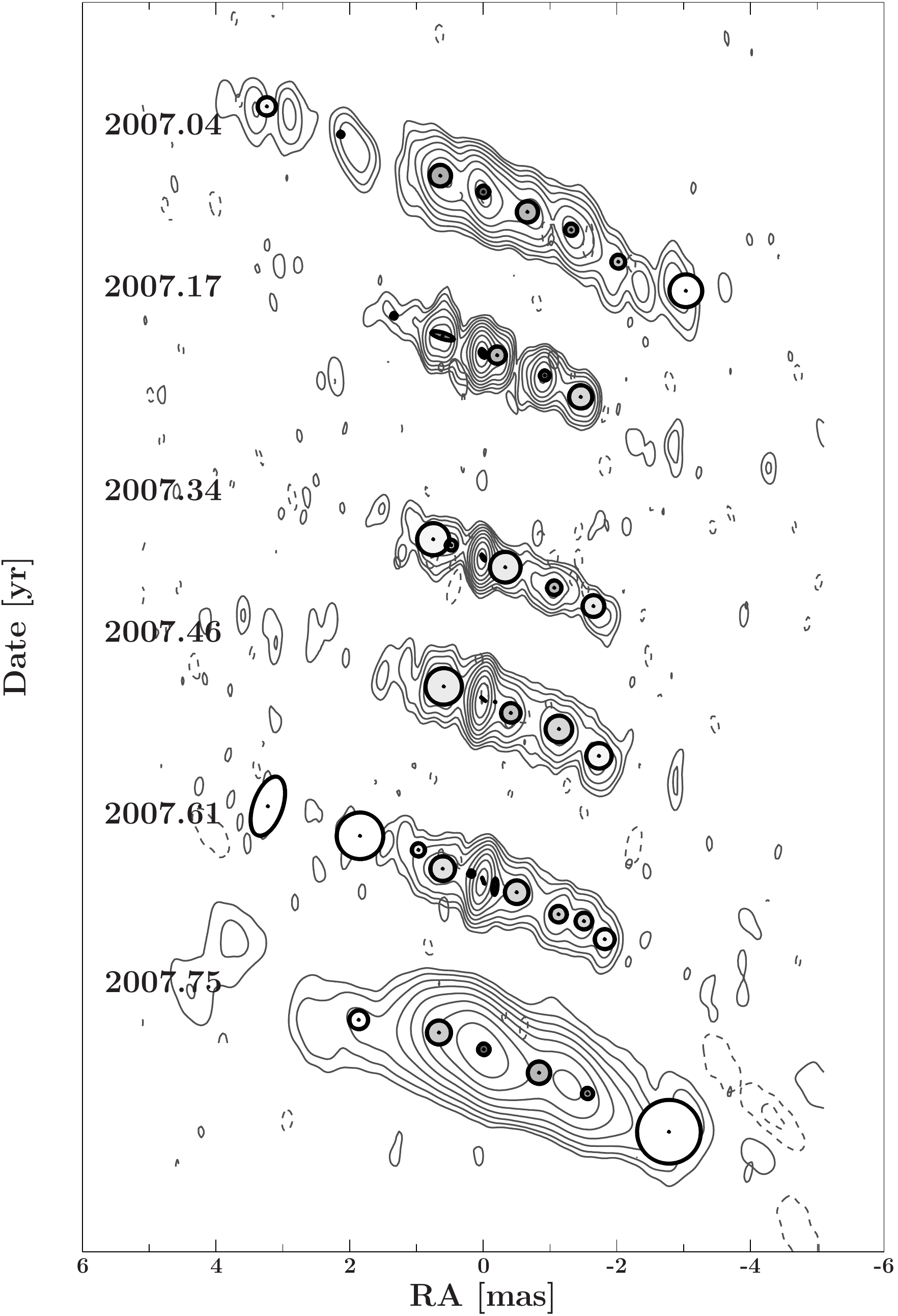}
  \caption{\textsc{clean} contour maps with over-plotted model components for the first and second block at 43\,GHz. The date of observation is printed left of each image. The brightness temperatures of the components are represented by a filling factor, with brigher components corresponding to lower brightness temperatures. The contours start at 3 times the noise level and increase logarithmically by factors of 2.}
  \label{fig:TimeEvo1}
\end{figure*}

\begin{figure}[!h]
\centering
  \includegraphics[width=0.98\linewidth]{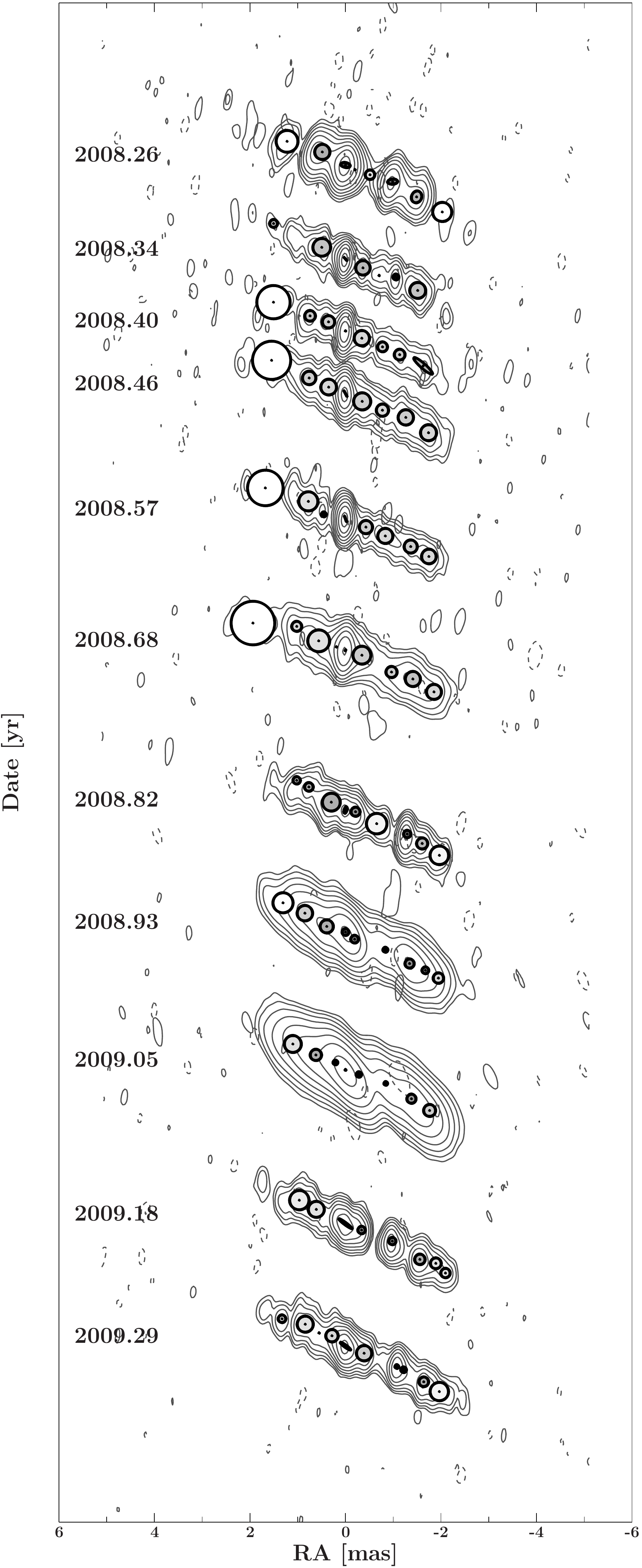}
  \caption{\textsc{clean} contour maps with over-plotted model for the third block at 43\,GHz. The date of observation is printed left of each image. The brightness temperatures of the components are represented by a filling factor, with brigher components corresponding to lower brightness temperatures. The contours start at 3 times the noise level and increase logarithmically by factors of 2.}
  \label{fig:TimeEvo2}
\end{figure}


\begin{figure*}[!ht]
\centering
  \includegraphics[width=0.43\linewidth]{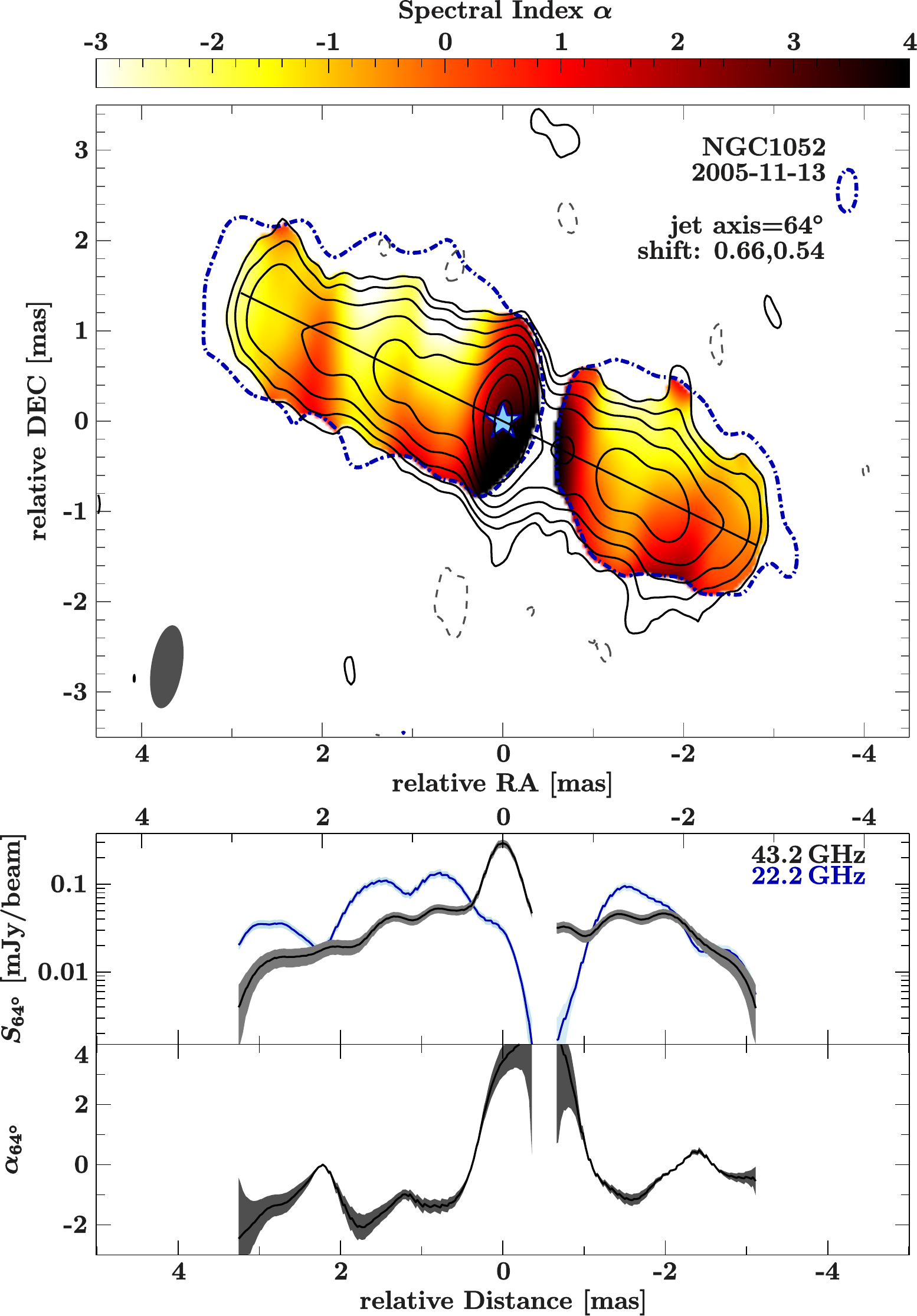}\hspace{0.04\linewidth}
  \includegraphics[width=0.43\linewidth]{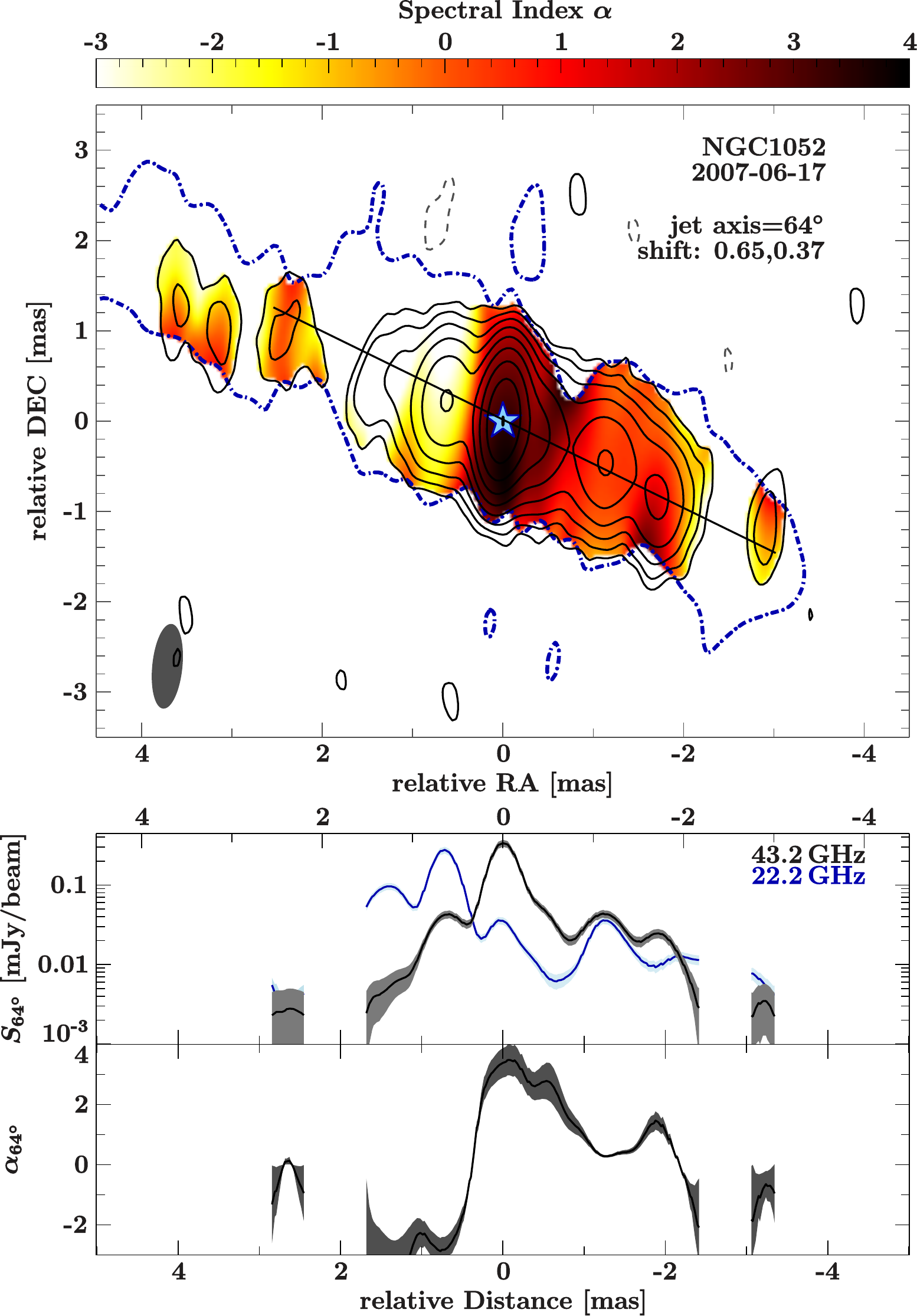}\vspace{0.5cm}
 
 \includegraphics[width=0.43\linewidth]{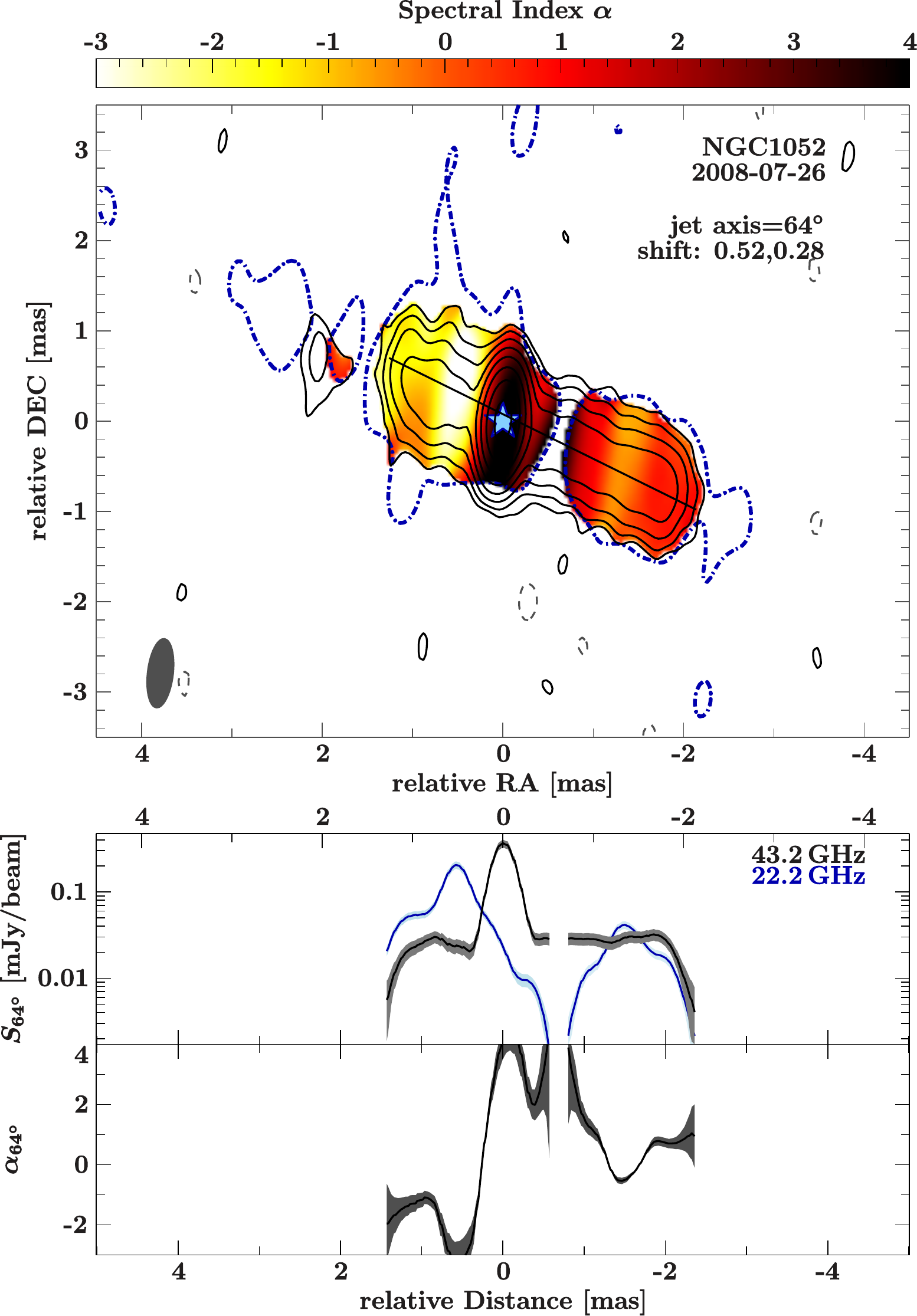}
 \caption{ Spectral index maps between 22\,GHz (dotted-dashed blue contour) and 43\,GHz (solid black contours) for three selected observations after alignment based on optically thin features (compare Fig.~\ref{fig:spix_blocks_examples}). The restoring beam for each epoch is plotted in the lower left corner. A cut along the direction of the jets ($64^\circ$) results in the lower panels, showing the spectral index (bottom panels) and the flux density (middle panel) along the jet axis. The blue star marks the assumed location of the dynamical center.}
 \label{fig:spixmaps_examples_extended}
\end{figure*}

\begin{table}[!h]
 \caption{Values of Reduce Chi square from Difmap for Gaussian Modelfitting}
 \label{fig:chi_square}
 \begin{tabular}{lc||lc}\hline\hline
	Epoch & $\chi^2$ & Epoch & $\chi^2$	\\\hline
	BR099A&  1.0642407 & BR120E&  1.0231979\\       
	BR099C&  0.9321564 & BR120F&  0.6463079\\       
	BR099D&  1.0673188 & BR120G&  1.2473867\\       
	BR099E&  1.0715107 & BR120H&  0.8063296\\       
	BR099F&  0.9859955 & BR120I&  1.0824018\\       
	BR099G&  1.1014646 & BR130A&  1.0717157\\       
	BR099H&  1.0176922 & BR130B&  1.0158143\\       
	BR099I&  0.9272598 & BR130C&  0.9345456\\       
	BR119A&  0.9957912 & BR130D&  0.9786728\\       
	BR119B&  0.9560460 & BR130E&  0.9293561\\       
	BR120A&  0.8111044 & BR130F&  0.8472571\\       
	BR120B&  0.8067314 & BR130G&  0.7743667\\       
	BR120C&  0.9757917 & BR130H&  0.9449142\\       
	BR120D&  0.9645245 & BR130I&  0.9357334\\\hline 
 \end{tabular}
\end{table}

\begin{table}
 \setlength{\tabcolsep}{1.5pt}
\tiny
\centering
\caption{Parameters for all components of epochs BR099A, BR099C, BR099D, BR099E, BR099F (from top to bottom)}
\label{tab:timeevo_fits_single1}
\input{model_paras_epoch0.tex}
\input{model_paras_epoch1.tex}
\input{model_paras_epoch2.tex}
\input{model_paras_epoch3.tex}
\input{model_paras_epoch4.tex}
\end{table}

\begin{table}
 \setlength{\tabcolsep}{1.5pt}
\tiny
\centering
\caption{Parameters for all components of epoch BR099G, BR099I, BR119A, BR119B and BR120A,  (from top to bottom).  Delta-like components with an unphysical high brightness temperature larger than $10^{18}\,$K have been excluded for each analysis based on brighness temperatures.}
\label{tab:timeevo_fits_single2}

\input{model_paras_epoch5.tex}
\input{model_paras_epoch6.tex}
\input{model_paras_epoch7.tex}
\input{model_paras_epoch8.tex}
\input{model_paras_epoch9.tex}
\end{table}

\begin{table}
 \setlength{\tabcolsep}{1.5pt}
\tiny
\centering
\caption{Parameters for all components of epoch BR120B, BR120C, BR120D, BR120E, BR120F, BR120H and BR120I (from top to bottom). Delta-like components with an unphysical high brightness temperature larger than $10^{18}\,$K have been excluded for each analysis based on brighness temperatures.}
\label{tab:timeevo_fits_single3}\vspace{0.1cm}
\input{model_paras_epoch10.tex}
\input{model_paras_epoch11.tex}
\input{model_paras_epoch12.tex}
\input{model_paras_epoch13.tex}
\input{model_paras_epoch14.tex}
\input{model_paras_epoch15.tex}
\input{model_paras_epoch16.tex}
\end{table}

\begin{table}
 \setlength{\tabcolsep}{1.5pt}
\tiny
\centering
\caption{Parameters for all components of epoch BR130A, BR130B, BR130C, BR130D, BR130E and BR130F (from top to bottom). Delta-like components with an unphysical high brightness temperature larger than $10^{18}\,$K have been excluded for each analysis based on brighness temperatures.}
\label{tab:timeevo_fits_single4}\vspace{0.1cm}
\input{model_paras_epoch17.tex}
\input{model_paras_epoch18.tex}
\input{model_paras_epoch19.tex}
\input{model_paras_epoch20.tex}
\input{model_paras_epoch21.tex}
\input{model_paras_epoch22.tex}
\end{table}

\begin{table}
 \setlength{\tabcolsep}{1.5pt}
\tiny
\centering
\caption{Parameters for all components of epoch BR130G, BR130H and BR130I (from top to bottom). Delta-like components with an unphysical high brightness temperature larger than $10^{18}\,$K have been excluded for each analysis based on brighness temperatures.}
\label{tab:timeevo_fits_single5}\vspace{0.1cm}
\input{model_paras_epoch23.tex}
\input{model_paras_epoch24.tex}
\input{model_paras_epoch25.tex}
\end{table}
\end{appendix}
\end{document}

%% file: model_paras_epoch0.tex
\begin{tabular}{ccccccc}
 \hline
 \hline
 Component & $S_\mathrm{tot}$ [Jy] & Distance [mas] & P.A. $[^\circ]$ & Major [mas] & Ratio & $\log\, T_\mathrm{b} [K]$ \\
 \hline
 wj1 & 0.07 & -2.36 & -- & 0.36 & 1.00 & 8.56 \\
 wj2 & 0.09 & -1.81 & -- & 0.28 & 1.00 & 8.89 \\
 wj4 & 0.08 & -1.16 & -- & 0.21 & 1.00 & 9.07 \\
 -- & 0.05 & -0.61 & 23.84 & 0.50 & 0.27 & 8.73 \\
 core & 0.38 & 0.00 & 43.83 & 0.15 & 0.00 & -- \\
 ej3 & 0.08 & 0.59 & -- & 0.40 & 1.00 & 8.50 \\
 ej2 & 0.07 & 1.12 & -- & 0.30 & 1.00 & 8.68 \\
 ej1 & 0.04 & 1.66 & -- & 0.33 & 1.00 & 8.38 \\
 -- & 0.06 & 2.28 & 40.86 & 0.65 & 0.58 & 8.24 \\
 \hline
\end{tabular}

%% file: model_paras_epoch1.tex
\begin{tabular}{ccccccc}
 \hline
 \hline
 Component & $S_\mathrm{tot}$ [Jy] & Distance [mas] & P.A. $[^\circ]$ & Major [mas] & Ratio & $\log\, T_\mathrm{b} [K]$ \\
 \hline
 wj1 & 0.05 & -2.80 & -- & 0.47 & 1.00 & 8.16 \\
 wj2 & 0.12 & -2.18 & -- & 0.34 & 1.00 & 8.85 \\
 wj3 & 0.08 & -1.79 & -- & 0.15 & 1.00 & 9.36 \\
 wj4 & 0.09 & -1.41 & -- & 0.36 & 1.00 & 8.68 \\
 wj5 & 0.11 & -0.47 & -- & 0.30 & 1.00 & 8.92 \\
 core & 0.43 & 0.00 & 5.50 & 0.13 & 0.39 & 10.63 \\
 ej4 & 0.07 & 0.48 & -- & 0.07 & 1.00 & 9.96 \\
 ej3 & 0.15 & 0.82 & -- & 0.28 & 1.00 & 9.10 \\
 ej2 & 0.05 & 1.53 & -- & 0.50 & 1.00 & 8.12 \\
 ej1 & 0.03 & 2.19 & -- & 0.18 & 1.00 & 8.71 \\
 -- & 0.03 & 2.58 & -- & 0.22 & 1.00 & 8.58 \\
 \hline
\end{tabular}

%% file: model_paras_epoch2.tex
\begin{tabular}{ccccccc}
 \hline
 \hline
 Component & $S_\mathrm{tot}$ [Jy] & Distance [mas] & P.A. $[^\circ]$ & Major [mas] & Ratio & $\log\, T_\mathrm{b} [K]$ \\
 \hline
 wj1 & 0.02 & -3.13 & -- & 0.21 & 1.00 & 8.38 \\
 wj2 & 0.08 & -2.40 & -- & 0.39 & 1.00 & 8.53 \\
 wj3 & 0.08 & -1.92 & -- & 0.23 & 1.00 & 9.02 \\
 wj4 & 0.06 & -1.49 & -- & 0.32 & 1.00 & 8.62 \\
 wj5 & 0.05 & -0.78 & -- & 0.36 & 1.00 & 8.46 \\
 wj6 & 0.09 & -0.31 & -- & 0.31 & 1.00 & 8.82 \\
 core & 0.31 & 0.00 & 18.66 & 0.10 & 0.00 & -- \\
 ej5 & 0.12 & 0.27 & -- & 0.41 & 1.00 & 8.66 \\
 ej4 & 0.08 & 0.78 & -- & 0.23 & 1.00 & 9.04 \\
 ej3 & 0.06 & 1.24 & -- & 0.37 & 1.00 & 8.45 \\
 ej2 & 0.05 & 1.87 & -- & 0.52 & 1.00 & 8.04 \\
 ej1 & 0.05 & 2.50 & -- & 0.44 & 1.00 & 8.23 \\
 \hline
\end{tabular}

%% file: model_paras_epoch3.tex
\begin{tabular}{ccccccc}
 \hline
 \hline
 Component & $S_\mathrm{tot}$ [Jy] & Distance [mas] & P.A. $[^\circ]$ & Major [mas] & Ratio & $\log\, T_\mathrm{b} [K]$ \\
 \hline
 wj1 & 0.02 & -3.23 & -- & 0.29 & 1.00 & 8.23 \\
 wj2 & 0.09 & -2.51 & -- & 0.32 & 1.00 & 8.77 \\
 wj3 & 0.16 & -1.99 & -- & 0.29 & 1.00 & 9.11 \\
 wj4 & 0.09 & -1.57 & -- & 0.31 & 1.00 & 8.82 \\
 wj5 & 0.07 & -0.82 & -- & 0.24 & 1.00 & 8.87 \\
 wj6 & 0.19 & -0.34 & -- & 0.35 & 1.00 & 9.02 \\
 core & 0.48 & 0.00 & 12.03 & 0.18 & 0.00 & -- \\
 ej5 & 0.15 & 0.31 & -- & 0.37 & 1.00 & 8.87 \\
 ej4 & 0.14 & 0.81 & -- & 0.27 & 1.00 & 9.11 \\
 ej3 & 0.09 & 1.35 & -- & 0.41 & 1.00 & 8.54 \\
 ej2 & 0.08 & 2.07 & -- & 0.43 & 1.00 & 8.42 \\
 ej1 & 0.04 & 2.57 & -- & 0.29 & 1.00 & 8.47 \\
 \hline
\end{tabular}

%% file: model_paras_epoch4.tex
\begin{tabular}{ccccccc}
 \hline
 \hline
 Component & $S_\mathrm{tot}$ [Jy] & Distance [mas] & P.A. $[^\circ]$ & Major [mas] & Ratio & $\log\, T_\mathrm{b} [K]$ \\
 \hline
 wj2 & 0.08 & -2.67 & -- & 0.47 & 1.00 & 8.37 \\
 wj3 & 0.11 & -2.07 & -- & 0.32 & 1.00 & 8.85 \\
 wj4 & 0.06 & -1.59 & -- & 0.18 & 1.00 & 9.07 \\
 wj5 & 0.06 & -1.04 & -- & 0.23 & 1.00 & 8.91 \\
 wj6 & 0.10 & -0.50 & -- & 0.15 & 1.00 & 9.51 \\
 core & 0.53 & 0.00 & -- & 0.08 & 1.00 & 10.74 \\
 ej5 & 0.14 & 0.58 & -- & 0.42 & 1.00 & 8.72 \\
 ej4 & 0.08 & 1.19 & -- & 0.24 & 1.00 & 8.95 \\
 ej3 & 0.07 & 1.74 & -- & 0.32 & 1.00 & 8.66 \\
 ej2 & 0.07 & 2.32 & -- & 0.43 & 1.00 & 8.38 \\
 ej1 & 0.04 & 2.92 & -- & 0.37 & 1.00 & 8.34 \\
 \hline
\end{tabular}

%% file: model_paras_epoch5.tex
\begin{tabular}{ccccccc}
 \hline
 \hline
 Component & $S_\mathrm{tot}$ [Jy] & Distance [mas] & P.A. $[^\circ]$ & Major [mas] & Ratio & $\log\, T_\mathrm{b} [K]$ \\
 \hline
 wj2 & 0.02 & -2.88 & -- & 0.17 & 1.00 & 8.61 \\
 wj3 & 0.07 & -2.17 & -- & 0.38 & 1.00 & 8.52 \\
 wj4 & 0.05 & -1.71 & -- & 0.25 & 1.00 & 8.71 \\
 wj5 & 0.05 & -1.33 & -- & 0.27 & 1.00 & 8.64 \\
 wj6 & 0.09 & -0.55 & -- & 0.52 & 1.00 & 8.35 \\
 core & 0.30 & 0.00 & -- & 0.09 & 1.00 & 10.40 \\
 ej6 & 0.04 & 0.50 & -- & 0.11 & 1.00 & 9.37 \\
 ej5 & 0.08 & 0.82 & -- & 0.35 & 1.00 & 8.64 \\
 ej4 & 0.05 & 1.33 & -- & 0.25 & 1.00 & 8.76 \\
 ej3 & 0.05 & 1.96 & -- & 0.55 & 1.00 & 8.03 \\
 ej2 & 0.04 & 2.73 & -- & 0.58 & 1.00 & 7.87 \\
 \hline
\end{tabular}

%% file: model_paras_epoch6.tex
\begin{tabular}{ccccccc}
 \hline
 \hline
 Component & $S_\mathrm{tot}$ [Jy] & Distance [mas] & P.A. $[^\circ]$ & Major [mas] & Ratio & $\log\, T_\mathrm{b} [K]$ \\
 \hline
 wj2 & 0.02 & -3.16 & -- & 0.47 & 1.00 & 7.88 \\
 wj3 & 0.07 & -2.22 & -- & 0.42 & 1.00 & 8.41 \\
 wj5 & 0.06 & -1.66 & -- & 0.28 & 1.00 & 8.70 \\
 wj6 & 0.05 & -0.83 & -- & 0.30 & 1.00 & 8.62 \\
 wj7 & 0.03 & -0.48 & -- & 0.00 & 1.00 & -- \\
 core & 0.36 & 0.00 & -- & 0.07 & 1.00 & 10.68 \\
 ej7 & 0.04 & 0.46 & -- & 0.00 & 1.00 & -- \\
 ej6 & 0.05 & 0.77 & -- & 0.16 & 1.00 & 9.15 \\
 ej5 & 0.02 & 1.26 & -- & 0.00 & 1.00 & 20.05 \\
 ej4 & 0.05 & 1.62 & -- & 0.27 & 1.00 & 8.64 \\
 ej3 & 0.02 & 2.14 & -- & 0.19 & 1.00 & 8.63 \\
 ej2 & 0.03 & 2.93 & -- & 0.45 & 1.00 & 7.99 \\
 \hline
\end{tabular}

%% file: model_paras_epoch7.tex
\begin{tabular}{ccccccc}
 \hline
 \hline
 Component & $S_\mathrm{tot}$ [Jy] & Distance [mas] & P.A. $[^\circ]$ & Major [mas] & Ratio & $\log\, T_\mathrm{b} [K]$ \\
 \hline
 wj3 & 0.05 & -2.41 & -- & 0.57 & 1.00 & 8.03 \\
 wj5 & 0.06 & -1.80 & -- & 0.36 & 1.00 & 8.47 \\
 wj6 & 0.05 & -1.12 & -- & 0.23 & 1.00 & 8.81 \\
 wj7 & 0.06 & -0.58 & -- & 0.48 & 1.00 & 8.24 \\
 core & 0.44 & 0.00 & 40.59 & 0.09 & 0.00 & 18.64 \\
 ej7 & 0.06 & 0.69 & -- & 0.36 & 1.00 & 8.48 \\
 ej6 & 0.04 & 1.16 & -- & 0.17 & 1.00 & 8.94 \\
 ej5 & 0.03 & 1.60 & -- & 0.33 & 1.00 & 8.29 \\
 ej4 & 0.04 & 2.03 & -- & 0.38 & 1.00 & 8.24 \\
 ej2 & 0.04 & 3.22 & -- & 0.75 & 1.00 & 7.65 \\
 \hline
\end{tabular}

%% file: model_paras_epoch8.tex
\begin{tabular}{ccccccc}
 \hline
 \hline
 Component & $S_\mathrm{tot}$ [Jy] & Distance [mas] & P.A. $[^\circ]$ & Major [mas] & Ratio & $\log\, T_\mathrm{b} [K]$ \\
 \hline
 wj2 & 0.02 & -3.60 & -- & 0.66 & 1.00 & 7.44 \\
 wj3 & 0.06 & -2.49 & -- & 0.52 & 1.00 & 8.14 \\
 wj5 & 0.06 & -1.91 & -- & 0.36 & 1.00 & 8.50 \\
 wj6 & 0.12 & -1.09 & -- & 0.43 & 1.00 & 8.64 \\
 wj7 & 0.03 & -0.80 & -- & 0.13 & 1.00 & 9.02 \\
 -- & 0.09 & -0.28 & -- & 0.34 & 1.00 & 8.71 \\
 core & 0.42 & 0.00 & 12.03 & 0.14 & 0.00 & -- \\
 -- & 0.04 & 0.22 & -- & 0.05 & 1.00 & 10.03 \\
 -- & 0.04 & 0.44 & -- & 0.08 & 1.00 & 9.57 \\
 ej7 & 0.05 & 0.79 & -- & 0.29 & 1.00 & 8.62 \\
 ej6 & 0.06 & 1.23 & -- & 0.31 & 1.00 & 8.60 \\
 ej5 & 0.06 & 1.83 & -- & 0.49 & 1.00 & 8.23 \\
 ej3 & 0.06 & 2.82 & -- & 0.79 & 1.00 & 7.79 \\
 ej1 & 0.02 & 3.91 & -- & 0.50 & 1.00 & 7.68 \\
 \hline
\end{tabular}

%% file: model_paras_epoch9.tex
\begin{tabular}{ccccccc}
 \hline
 \hline
 Component & $S_\mathrm{tot}$ [Jy] & Distance [mas] & P.A. $[^\circ]$ & Major [mas] & Ratio & $\log\, T_\mathrm{b} [K]$ \\
 \hline
 -- & 0.02 & -3.37 & -- & 0.49 & 1.00 & 7.75 \\
 -- & 0.03 & -2.27 & -- & 0.21 & 1.00 & 8.72 \\
 wj8 & 0.09 & -1.43 & -- & 0.18 & 1.00 & 9.25 \\
 wj9 & 0.13 & -0.72 & -- & 0.32 & 1.00 & 8.93 \\
 core & 0.32 & 0.00 & -- & 0.18 & 1.00 & 9.81 \\
 -- & 0.14 & 0.69 & -- & 0.32 & 1.00 & 8.95 \\
 -- & 0.02 & 2.30 & -- & 0.08 & 1.00 & 9.34 \\
 -- & 0.02 & 3.48 & -- & 0.27 & 1.00 & 8.14 \\
 \hline
\end{tabular}

%% file: model_paras_epoch10.tex
\begin{tabular}{ccccccc}
 \hline
 \hline
 Component & $S_\mathrm{tot}$ [Jy] & Distance [mas] & P.A. $[^\circ]$ & Major [mas] & Ratio & $\log\, T_\mathrm{b} [K]$ \\
 \hline
 wj8 & 0.05 & -1.60 & -- & 0.35 & 1.00 & 8.48 \\
 wj9 & 0.11 & -0.98 & -- & 0.15 & 1.00 & 9.55 \\
 wj10 & 0.07 & -0.21 & -- & 0.26 & 1.00 & 8.87 \\
 core & 0.28 & 0.00 & 26.09 & 0.14 & 0.62 & 10.21 \\
 ej9 & 0.14 & 0.66 & 73.04 & 0.35 & 0.27 & 9.44 \\
 ej8 & 0.01 & 1.45 & -- & 0.08 & 1.00 & 8.93 \\
 \hline
\end{tabular}

%% file: model_paras_epoch11.tex
\begin{tabular}{ccccccc}
 \hline
 \hline
 Component & $S_\mathrm{tot}$ [Jy] & Distance [mas] & P.A. $[^\circ]$ & Major [mas] & Ratio & $\log\, T_\mathrm{b} [K]$ \\
 \hline
 wj8 & 0.03 & -1.80 & -- & 0.32 & 1.00 & 8.27 \\
 wj9 & 0.06 & -1.16 & -- & 0.22 & 1.00 & 8.93 \\
 wj10 & 0.07 & -0.36 & -- & 0.46 & 1.00 & 8.32 \\
 core & 0.27 & 0.00 & 28.12 & 0.10 & 0.28 & 10.83 \\
 ej10 & 0.02 & 0.51 & -- & 0.16 & 1.00 & 8.76 \\
 ej9 & 0.05 & 0.80 & -- & 0.48 & 1.00 & 8.12 \\
 \hline
\end{tabular}

%% file: model_paras_epoch12.tex
\begin{tabular}{ccccccc}
 \hline
 \hline
 Component & $S_\mathrm{tot}$ [Jy] & Distance [mas] & P.A. $[^\circ]$ & Major [mas] & Ratio & $\log\, T_\mathrm{b} [K]$ \\
 \hline
 wj8 & 0.04 & -1.93 & -- & 0.38 & 1.00 & 8.23 \\
 wj9 & 0.08 & -1.22 & -- & 0.40 & 1.00 & 8.51 \\
 wj10 & 0.07 & -0.46 & -- & 0.29 & 1.00 & 8.71 \\
 -- & 0.03 & -0.18 & -- & 0.00 & 1.00 & 20.59 \\
 core & 0.33 & 0.00 & 49.60 & 0.08 & 0.00 & -- \\
 ej10 & 0.09 & 0.62 & -- & 0.55 & 1.00 & 8.31 \\
 \hline
\end{tabular}

%% file: model_paras_epoch13.tex
\begin{tabular}{ccccccc}
 \hline
 \hline
 Component & $S_\mathrm{tot}$ [Jy] & Distance [mas] & P.A. $[^\circ]$ & Major [mas] & Ratio & $\log\, T_\mathrm{b} [K]$ \\
 \hline
 wj8 & 0.02 & -2.02 & -- & 0.30 & 1.00 & 8.15 \\
 -- & 0.03 & -1.62 & -- & 0.25 & 1.00 & 8.52 \\
 wj9 & 0.04 & -1.23 & -- & 0.25 & 1.00 & 8.69 \\
 wj10 & 0.06 & -0.53 & -- & 0.35 & 1.00 & 8.51 \\
 -- & 0.04 & -0.20 & -4.58 & 0.24 & 0.30 & 9.12 \\
 core & 0.27 & 0.00 & 20.67 & 0.09 & 0.00 & -- \\
 -- & 0.02 & 0.21 & -- & 0.09 & 1.00 & 9.22 \\
 ej10 & 0.05 & 0.63 & -- & 0.37 & 1.00 & 8.40 \\
 ej9 & 0.01 & 1.07 & -- & 0.20 & 1.00 & 8.30 \\
 ej8 & 0.01 & 1.96 & -- & 0.70 & 1.00 & 7.24 \\
 -- & 0.01 & 3.41 & -19.34 & 0.93 & 0.48 & 7.32 \\
 \hline
\end{tabular}

%% file: model_paras_epoch14.tex
\begin{tabular}{ccccccc}
 \hline
 \hline
 Component & $S_\mathrm{tot}$ [Jy] & Distance [mas] & P.A. $[^\circ]$ & Major [mas] & Ratio & $\log\, T_\mathrm{b} [K]$ \\
 \hline
 -- & 0.02 & -3.03 & -- & 0.95 & 1.00 & 7.11 \\
 wj9 & 0.08 & -1.69 & -- & 0.17 & 1.00 & 9.28 \\
 wj10 & 0.12 & -0.90 & -- & 0.34 & 1.00 & 8.85 \\
 core & 0.33 & 0.00 & -- & 0.17 & 1.00 & 9.88 \\
 ej10 & 0.09 & 0.72 & -- & 0.36 & 1.00 & 8.65 \\
 ej8 & 0.01 & 1.92 & -- & 0.28 & 1.00 & 7.95 \\
 \hline
\end{tabular}

%% file: model_paras_epoch15.tex
\begin{tabular}{ccccccc}
 \hline
 \hline
 Component & $S_\mathrm{tot}$ [Jy] & Distance [mas] & P.A. $[^\circ]$ & Major [mas] & Ratio & $\log\, T_\mathrm{b} [K]$ \\
 \hline
 -- & 0.01 & -2.25 & -- & 0.40 & 1.00 & 7.63 \\
 wj11 & 0.06 & -1.63 & -22.40 & 0.25 & 0.84 & 8.87 \\
 wj12 & 0.14 & -1.05 & -85.65 & 0.21 & 0.51 & 9.62 \\
 wj13 & 0.02 & -0.55 & -- & 0.20 & 1.00 & 8.43 \\
 wj14 & 0.01 & -0.22 & -- & 0.00 & 1.00 & 23.31 \\
 core & 0.33 & 0.00 & 80.64 & 0.20 & 0.52 & 10.04 \\
 ej12 & 0.14 & 0.56 & -- & 0.32 & 1.00 & 8.97 \\
 ej11 & 0.03 & 1.32 & -- & 0.46 & 1.00 & 7.91 \\
 \hline
\end{tabular}

%% file: model_paras_epoch16.tex
\begin{tabular}{ccccccc}
 \hline
 \hline
 Component & $S_\mathrm{tot}$ [Jy] & Distance [mas] & P.A. $[^\circ]$ & Major [mas] & Ratio & $\log\, T_\mathrm{b} [K]$ \\
 \hline
 wj11 & 0.05 & -1.78 & 48.69 & 0.51 & 0.20 & 8.83 \\
 wj12 & 0.04 & -1.24 & -- & 0.22 & 1.00 & 8.68 \\
 wj13 & 0.03 & -0.84 & -- & 0.21 & 1.00 & 8.64 \\
 wj14 & 0.05 & -0.37 & -- & 0.32 & 1.00 & 8.54 \\
 core & 0.26 & 0.00 & -- & 0.00 & 1.00 & 22.56 \\
 ej13 & 0.05 & 0.40 & -- & 0.26 & 1.00 & 8.67 \\
 ej12 & 0.04 & 0.81 & -- & 0.22 & 1.00 & 8.72 \\
 ej11 & 0.03 & 1.62 & -- & 0.70 & 1.00 & 7.55 \\
 \hline
\end{tabular}

%% file: model_paras_epoch17.tex
\begin{tabular}{ccccccc}
 \hline
 \hline
 Component & $S_\mathrm{tot}$ [Jy] & Distance [mas] & P.A. $[^\circ]$ & Major [mas] & Ratio & $\log\, T_\mathrm{b} [K]$ \\
 \hline
 wj11 & 0.10 & -1.65 & -- & 0.33 & 1.00 & 8.77 \\
 wj12 & 0.06 & -1.12 & -- & 0.12 & 1.00 & 9.42 \\
 wj13 & 0.02 & -0.78 & -- & 0.00 & 1.00 & 21.76 \\
 wj14 & 0.10 & -0.41 & -- & 0.30 & 1.00 & 8.86 \\
 core & 0.29 & 0.00 & 40.50 & 0.08 & 0.00 & -- \\
 ej12 & 0.12 & 0.55 & -- & 0.37 & 1.00 & 8.77 \\
 ej11 & 0.02 & 1.68 & -- & 0.16 & 1.00 & 8.73 \\
 \hline
\end{tabular}

%% file: model_paras_epoch18.tex
\begin{tabular}{ccccccc}
 \hline
 \hline
 Component & $S_\mathrm{tot}$ [Jy] & Distance [mas] & P.A. $[^\circ]$ & Major [mas] & Ratio & $\log\, T_\mathrm{b} [K]$ \\
 \hline
 wj11 & 0.06 & -1.92 & -- & 0.34 & 1.00 & 8.55 \\
 wj12 & 0.07 & -1.36 & -- & 0.32 & 1.00 & 8.65 \\
 wj13 & 0.03 & -0.85 & -- & 0.26 & 1.00 & 8.53 \\
 wj14 & 0.08 & -0.39 & -- & 0.36 & 1.00 & 8.61 \\
 core & 0.28 & 0.00 & 27.61 & 0.13 & 0.00 & -- \\
 ej13 & 0.06 & 0.38 & -- & 0.33 & 1.00 & 8.56 \\
 ej12 & 0.05 & 0.83 & -- & 0.28 & 1.00 & 8.60 \\
 ej11 & 0.02 & 1.70 & -- & 0.80 & 1.00 & 7.39 \\
 \hline
\end{tabular}

%% file: model_paras_epoch19.tex
\begin{tabular}{ccccccc}
 \hline
 \hline
 Component & $S_\mathrm{tot}$ [Jy] & Distance [mas] & P.A. $[^\circ]$ & Major [mas] & Ratio & $\log\, T_\mathrm{b} [K]$ \\
 \hline
 wj11 & 0.05 & -1.92 & -- & 0.31 & 1.00 & 8.55 \\
 wj12 & 0.05 & -1.48 & -- & 0.27 & 1.00 & 8.63 \\
 wj13 & 0.04 & -0.91 & -- & 0.32 & 1.00 & 8.47 \\
 wj14 & 0.05 & -0.46 & -- & 0.28 & 1.00 & 8.68 \\
 core & 0.37 & 0.00 & 19.85 & 0.12 & 0.00 & -- \\
 ej13 & 0.03 & 0.46 & -- & 0.10 & 1.00 & 9.23 \\
 ej12 & 0.07 & 0.86 & -- & 0.40 & 1.00 & 8.45 \\
 ej11 & 0.02 & 1.80 & -- & 0.74 & 1.00 & 7.34 \\
 \hline
\end{tabular}

%% file: model_paras_epoch20.tex
\begin{tabular}{ccccccc}
 \hline
 \hline
 Component & $S_\mathrm{tot}$ [Jy] & Distance [mas] & P.A. $[^\circ]$ & Major [mas] & Ratio & $\log\, T_\mathrm{b} [K]$ \\
 \hline
 wj11 & 0.05 & -2.05 & -- & 0.31 & 1.00 & 8.49 \\
 wj12 & 0.08 & -1.53 & -- & 0.32 & 1.00 & 8.71 \\
 wj13 & 0.04 & -1.07 & -- & 0.23 & 1.00 & 8.69 \\
 wj14 & 0.09 & -0.36 & -- & 0.37 & 1.00 & 8.62 \\
 core & 0.31 & 0.00 & -24.93 & 0.02 & 0.00 & -- \\
 ej14 & 0.02 & 0.20 & -- & 0.00 & 1.00 & 21.95 \\
 ej13 & 0.08 & 0.60 & -- & 0.46 & 1.00 & 8.42 \\
 ej12 & 0.02 & 1.14 & -- & 0.21 & 1.00 & 8.56 \\
 ej11 & 0.03 & 2.02 & -- & 0.91 & 1.00 & 7.34 \\
 \hline
\end{tabular}

%% file: model_paras_epoch21.tex
\begin{tabular}{ccccccc}
 \hline
 \hline
 Component & $S_\mathrm{tot}$ [Jy] & Distance [mas] & P.A. $[^\circ]$ & Major [mas] & Ratio & $\log\, T_\mathrm{b} [K]$ \\
 \hline
 wj11 & 0.03 & -2.18 & -- & 0.39 & 1.00 & 8.18 \\
 wj12 & 0.05 & -1.75 & -- & 0.23 & 1.00 & 8.83 \\
 wj13 & 0.14 & -1.39 & -- & 0.15 & 1.00 & 9.60 \\
 wj14 & 0.05 & -0.71 & -- & 0.43 & 1.00 & 8.23 \\
 wj15 & 0.13 & -0.21 & -- & 0.19 & 1.00 & 9.39 \\
 core & 0.09 & 0.00 & -14.80 & 0.14 & 0.42 & 9.85 \\
 ej14 & 0.22 & 0.34 & -- & 0.37 & 1.00 & 9.01 \\
 ej13 & 0.05 & 0.91 & -- & 0.18 & 1.00 & 9.02 \\
 ej12 & 0.03 & 1.20 & -- & 0.15 & 1.00 & 8.92 \\
 \hline
\end{tabular}

%% file: model_paras_epoch22.tex
\begin{tabular}{ccccccc}
 \hline
 \hline
 Component & $S_\mathrm{tot}$ [Jy] & Distance [mas] & P.A. $[^\circ]$ & Major [mas] & Ratio & $\log\, T_\mathrm{b} [K]$ \\
 \hline
 wj11 & 0.02 & -2.17 & -- & 0.22 & 1.00 & 8.50 \\
 wj12 & 0.04 & -1.85 & -- & 0.14 & 1.00 & 9.17 \\
 wj13 & 0.11 & -1.50 & -- & 0.19 & 1.00 & 9.30 \\
 wj14 & 0.05 & -0.91 & -- & 0.10 & 1.00 & 9.54 \\
 wj15 & 0.14 & -0.24 & -- & 0.16 & 1.00 & 9.56 \\
 core & 0.14 & 0.00 & -- & 0.14 & 1.00 & 9.68 \\
 ej14 & 0.12 & 0.41 & -- & 0.28 & 1.00 & 9.03 \\
 ej13 & 0.08 & 0.94 & -- & 0.32 & 1.00 & 8.70 \\
 ej12 & 0.02 & 1.44 & -- & 0.42 & 1.00 & 7.82 \\
 \hline
\end{tabular}

%% file: model_paras_epoch23.tex
\begin{tabular}{ccccccc}
 \hline
 \hline
 Component & $S_\mathrm{tot}$ [Jy] & Distance [mas] & P.A. $[^\circ]$ & Major [mas] & Ratio & $\log\, T_\mathrm{b} [K]$ \\
 \hline
 wj12 & 0.05 & -1.95 & -- & 0.25 & 1.00 & 8.70 \\
 wj13 & 0.09 & -1.51 & -- & 0.19 & 1.00 & 9.20 \\
 wj14 & 0.07 & -0.89 & -- & 0.07 & 1.00 & 9.93 \\
 wj15 & 0.08 & -0.30 & -- & 0.11 & 1.00 & 9.65 \\
 core & 0.16 & 0.00 & -- & 0.01 & 1.00 & 11.73 \\
 ej15 & 0.11 & 0.26 & -- & 0.10 & 1.00 & 9.90 \\
 ej14 & 0.08 & 0.70 & -- & 0.24 & 1.00 & 9.00 \\
 ej13 & 0.06 & 1.23 & -- & 0.36 & 1.00 & 8.46 \\
 \hline
\end{tabular}

%% file: model_paras_epoch24.tex
\begin{tabular}{ccccccc}
 \hline
 \hline
 Component & $S_\mathrm{tot}$ [Jy] & Distance [mas] & P.A. $[^\circ]$ & Major [mas] & Ratio & $\log\, T_\mathrm{b} [K]$ \\
 \hline
 wj11 & 0.01 & -2.33 & -- & 0.20 & 1.00 & 8.38 \\
 wj12 & 0.02 & -2.06 & -- & 0.23 & 1.00 & 8.42 \\
 wj13 & 0.08 & -1.73 & -- & 0.22 & 1.00 & 9.02 \\
 wj14 & 0.09 & -1.04 & -- & 0.15 & 1.00 & 9.41 \\
 wj15 & 0.06 & -0.36 & -- & 0.15 & 1.00 & 9.25 \\
 core & 0.28 & 0.00 & 53.49 & 0.34 & 0.15 & 10.04 \\
 ej14 & 0.05 & 0.68 & -- & 0.33 & 1.00 & 8.45 \\
 ej13 & 0.05 & 1.09 & -- & 0.41 & 1.00 & 8.34 \\
 \hline
\end{tabular}

%% file: model_paras_epoch25.tex
\begin{tabular}{ccccccc}
 \hline
 \hline
 Component & $S_\mathrm{tot}$ [Jy] & Distance [mas] & P.A. $[^\circ]$ & Major [mas] & Ratio & $\log\, T_\mathrm{b} [K]$ \\
 \hline
 wj12 & 0.04 & -2.18 & -- & 0.39 & 1.00 & 8.22 \\
 wj13 & 0.06 & -1.80 & -- & 0.20 & 1.00 & 8.97 \\
 -- & 0.02 & -1.31 & -- & 0.12 & 1.00 & 9.08 \\
 wj14 & 0.07 & -1.16 & -- & 0.08 & 1.00 & 9.91 \\
 wj15 & 0.07 & -0.42 & -- & 0.32 & 1.00 & 8.68 \\
 core & 0.28 & 0.00 & 54.02 & 0.26 & 0.19 & 10.15 \\
 -- & 0.04 & 0.36 & -- & 0.26 & 1.00 & 8.56 \\
 -- & 0.01 & 0.62 & -- & 0.00 & 1.00 & 21.90 \\
 ej14 & 0.05 & 0.96 & -- & 0.34 & 1.00 & 8.45 \\
 ej13 & 0.01 & 1.45 & -- & 0.16 & 1.00 & 8.57 \\
 \hline
\end{tabular}